\documentclass[sigconf,nonacm]{acmart}

\usepackage{tcolorbox}

\setcopyright{none}

\newcommand\vldbavailabilityurl{https://github.com/pminhtam/AV-SQL}

\usepackage{hyperref}

\usepackage{epstopdf}
\usepackage{graphics}
\DeclareGraphicsExtensions{.eps,.gz,.eps,.pdf,.png,.jpg}
\graphicspath{{./}{./figures/}}

\usepackage{multirow}
\usepackage{paralist}

\newcommand{\sstitle}[1]{\noindent\textbf{#1.\/}}

\usepackage{pifont}
\newcommand{\cmark}{\ding{51}} %
\newcommand{\xmark}{\ding{55}} %

 \def\Snospace~{\S{}}

\usepackage{xspace}
\newcommand{\toolname}{\texttt{AV-SQL}\xspace}

\usepackage[ruled,vlined]{algorithm2e}
\SetKwProg{Fn}{Function}{:}{}
\SetKwProg{Proc}{Procedure}{:}{}

\usepackage{makecell}

\usepackage{adjustbox}
\tcbuselibrary{listings,breakable}
\usepackage{longtable}

\begin{document}

\setlength{\belowdisplayskip}{3pt}
\setlength{\belowdisplayshortskip}{3pt}
\setlength{\abovedisplayskip}{3pt}
\setlength{\abovedisplayshortskip}{3pt}

\title{AV-SQL: Decomposing Complex Text-to-SQL Queries with Agentic Views}

\author[Minh Tam Pham et al.]{Minh Tam Pham$^1$, Trinh Pham$^1$, Tong Chen$^2$, Hongzhi Yin$^2$,\\Quoc Viet Hung Nguyen$^1$, Thanh Tam Nguyen$^1$}
\affiliation{
  \institution{
    $^1$Griffith University (Australia),
    $^2$The University of Queensland (Australia)
  }
  \country{\vspace{.8em}}
}

\begin{abstract}
  Text-to-SQL is the task of translating natural language queries into executable SQL for a given database, enabling non-expert users to access structured data without writing SQL manually. Despite rapid advances driven by large language models (LLMs), existing approaches still struggle with \textit{complex} queries in real-world settings, where database schemas are large and questions require multi-step reasoning over many interrelated tables. In such cases, providing the full schema often exceeds the context window, while one-shot generation frequently produces non-executable SQL due to syntax errors and incorrect schema linking. To address these challenges, we introduce \toolname, a framework that \textit{decomposes} complex Text-to-SQL into a pipeline of specialized LLM agents. Central to \toolname\ is the concept of \textit{agentic views}: agent-generated Common Table Expressions (CTEs) that encapsulate intermediate query logic and filter relevant schema elements from large schemas. \toolname\ operates in three stages: (1) a rewriter agent compresses and clarifies the input query; (2) a view generator agent processes schema chunks to produce agentic views; and (3) a planner, generator, and revisor agent collaboratively compose these views into the final SQL query. Extensive experiments show that \toolname achieves \(70.38\%\) execution accuracy on the challenging Spider 2.0 benchmark, outperforming state-of-the-art baselines, while remaining competitive on standard datasets with \(85.59\%\) on Spider, \(72.16\%\) on BIRD and \(63.78\%\) on KaggleDBQA.
\end{abstract}

\maketitle

\ifdefempty{\vldbavailabilityurl}{}{
  \vspace{.3cm}
  \begingroup\small\noindent\raggedright\textbf{Artifact Availability:}\\
  The source code is available at \url{\vldbavailabilityurl}.
  \endgroup
}

\section{Introduction}
\label{sec:intro}

\begin{figure}[!h]
  \centering
  \includegraphics[width=0.99\linewidth]{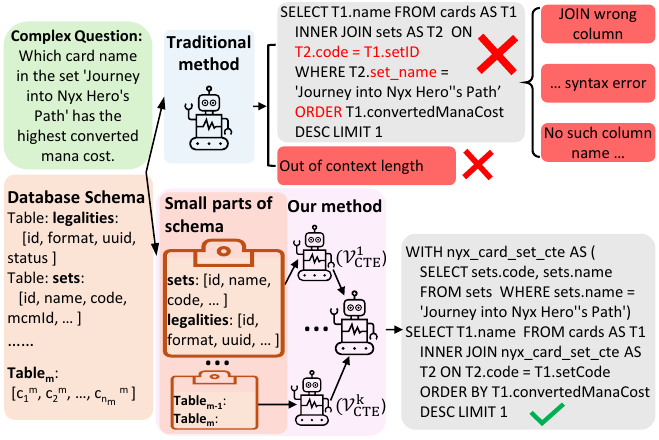}
  \vspace{-.5em}
  \caption{A motivating example from BIRD-dev (sample 501) illustrates the limitations of traditional Text-to-SQL on complex questions over large schemas. Providing the full schema in a single prompt can exceed the context window and increase the risk of syntax and logical errors. Our approach instead splits the schema into smaller chunks, generates and validates agent views (CTE queries: $\mathcal{V}^{(1)}_{\text{CTE}},$\dots $\mathcal{V}^{(k)}_{\text{CTE}}$), and then composes them into the final SQL, improving robustness under long-context constraints.}
  \label{fig:motivating_example}
  \vspace{-1em}
\end{figure}

Text-to-SQL translates a natural language question into an executable SQL query, enabling natural language interfaces to relational databases. Recent advances in large language models (LLMs) have improved semantic parsing and compositional SQL generation, especially on benchmarks such as Spider-1.0~\cite{yu-etal-2018-spider} and BIRD~\cite{li2024can}.
However, most Text-to-SQL systems~\cite{liu2025survey, pourreza2023dinsql, gao2023text, dong2023c3, li2024codes} implicitly assume that the full database schema can be provided in the prompt. While feasible for small schemas, this becomes impractical for real-world databases with thousands of tables and columns due to context-window limits, degraded attention in long prompts, and increased risk of schema-linking mistakes. Robustness is another barrier: even strong methods~\cite{talaei2024chess, li2024codes, wang2024mac, pourreza2023dinsql, luo2025natural} frequently generate SQL that fails to execute because of syntax or invalid references, and repairing a single complex query is difficult when errors are entangled. As illustrated in~\autoref{fig:motivating_example}, these issues arise when complex questions lead to diverse SQL errors that are difficult to fix.
\begin{example}
  \label{ex:motivating_example}
  The complex question in~\autoref{fig:motivating_example}, when handled by traditional Text-to-SQL, may produce a  SQL query with multiple intertwined errors, including incorrect joins (e.g., \texttt{T2.code=T1.setID}, where \texttt{T1.setID} is not a foreign key),, schema-linking mistakes (e.g., referencing the non-existent column \texttt{T2.set\_name}) , and syntax errors (e.g., omitting \texttt{BY} after \texttt{ORDER}).
  Because execution feedback is often limited to syntax errors and non-existent columns, it is difficult to identify and repair all of these errors holistically.
  Moreover, fixing  one error may expose or even introduce others, making it challenging to produce an executable query.
\end{example}

Spider-2.0~\cite{lei2024spider} makes these limitations explicit by introducing complex questions over large, intricate schemas that often cannot fit into an LLM's context window. Moreover, many gold queries rely on Common Table Expressions (CTEs) to organize multi-step logic, which improves readability for humans but increases the surface area for syntactic and logical errors in automatic generation. Consequently, one-shot generation with global schema prompting becomes unreliable at real-world scale.

To address these challenges, we propose \toolname, a multi-stage framework for robust Text-to-SQL.
Inspired by Chain-of-Agents long-context decomposition~\cite{zhang2024chain}, \toolname partitions the full schema into table-centric chunks and processes them sequentially.
However, unlike standard long-context summarization, Text-to-SQL requires exact schema grounding, so intermediate representations must preserve table and column names rather than compress them away.
To achieve this, \toolname treats chunk-level reasoning as schema filtering and represents intermediate decisions as executable CTEs. Specifically, \toolname consists of three stages: (1) \emph{Question Rewriting}, where a rewriter agent reformulates the question and external evidence into a concise, explicit form; (2) \emph{Agent View Generation}, where view generator agents examine schema chunks, produce CTE-based agent views, and validates them through execution with iterative repair; and (3) \emph{SQL Generation}, where planner, SQL generator, and revisor agents compose the validated agent views into the final executable SQL. As shown in~\autoref{fig:motivating_example}, this design reduces long-context pressure, makes schema filtering explicit, and improves executability by validating intermediate CTEs before final synthesis.
\begin{example}
  In~\autoref{fig:motivating_example}, \toolname\ partitions the schema into smaller chunks and lets each view generator focus on one chunk at a time. For a chunk containing the \texttt{sets} and \texttt{legalities} tables, the agent produces \texttt{nyx\_card\_set\_cte}, an executable CTE that captures the information needed for the question. Because this CTE only uses the \texttt{sets} table, the \texttt{legalities} table can be identified as irrelevant and excluded from later stages, which reduces context overload.  After iterative validation and repair to ensure executability and correctness,  this CTE can be directly incorporated into the final SQL query, lowering the risk of syntax and schema-linking errors.
\end{example}

Motivated by these design advantages, our main contributions are as follows:
\begin{itemize}
  \item \textbf{A novel multi-stage agentic Text-to-SQL framework.} \toolname introduces  a three-stage framework using specialized agents for question reformulation, CTE-based intermediate view construction, and final SQL synthesis.
  \item \textbf{CTE-based agent views with execution-guided repair.} Agent views are introduced as executable CTEs that make intermediate schema-filtering decisions explicit and verifiable, using  CTE-level execution feedback to iteratively repair errors before final SQL synthesis.
  \item \textbf{Strong empirical performance.} Extensive experiments show that \toolname\ outperforms state-of-the-art methods on large and complex real-world benchmarks, particularly Spider 2.0, while remaining competitive on standard benchmarks such as Spider, BIRD, and KaggleDBQA.
  \item \textbf{Zero-shot setting.} \toolname performs strongly in a fully zero-shot setting, requiring neither post-training nor in-context exemplars, and achieves strong results on Spider 2.0-Snow, Spider, BIRD, and KaggleDBQA.
  \item \textbf{Open-source release.} All code and artifacts will be released to support reproducibility and future research.
\end{itemize}
The rest of the paper is organized as follows: \autoref{sec:related} reviews related work; \autoref{sec:model} formally defines the problem and describes the overview of the \toolname\ framework; \autoref{sec:method} details the design of each agent and the overall pipeline; \autoref{sec:exp} presents experimental results and analysis; and \autoref{sec:conclusion} concludes with a summary.
\section{Related Work}
\label{sec:related}

\sstitle{Text-to-SQL benchmarks}
Early single-domain benchmarks include  Academic~\cite{li2014constructing}, MAS, IMDB, and YELP~\cite{yaghmazadeh2017sqlizer},  and Advising~\cite{finegan2018improving}. Later work broadened the scope: WikiSQL~\cite{zhong2017seq2sql} enabled large-scale table-based evaluation; Spider~\cite{yu-etal-2018-spider} introduced cross-domain multi-table SQL; BIRD~\cite{li2024can} covered 37 professional domains; and KaggleDBQA~\cite{lee2021kaggledbqa} targeted real-world web databases. Spider 2.0~\cite{lei2024spider} further raises the bar with complex queries over schemas with thousands of columns, making full-schema prompting impractical.

\sstitle{Text-to-SQL methods}
Text-to-SQL modeling has progressed through several stages.
Pre-LLM neural semantic parsers relied on structured decoding and explicit schema modeling. RAT-SQL~\cite{wang-etal-2020-rat} used relation-aware attention for schema linking; and RESDSQL~\cite{li2023resdsql} improved robustness via schema pruning. RASAT~\cite{qi-etal-2022-rasat}, and LGESQL~\cite{cao-etal-2021-lgesql} further refined graph-based schema-question modeling, while CatSQL~\cite{fu2023catsql} combined rule-based and neural approaches via a sketch-based framework.
With LLMs, early work~\cite{rajkumar2022evaluating, liu2023comprehensive} showed that zero- and few-shot prompting yields competitive performance without task-specific training. Methods such as DIN-SQL~\cite{pourreza2023dinsql}, MAC-SQL~\cite{wang2024mac}, C3SQL~\cite{dong2023c3}, DAIL-SQL~\cite{gao2023text}, DTS-SQL~\cite{pourreza2024dts}, and ZeroNL2SQL~\cite{fan2024combining} improved performance via task decomposition, demonstrations, and multi-model pipelines. Supervised fine-tuning approaches including CodeS~\cite{li2024codes} and XiYan-SQL~\cite{gao2024xiyan} strengthened open-source models, while hybrid and multi-candidate methods such as CHESS~\cite{talaei2024chess}, MCS-SQL~\cite{lee2025mcs}, and Super-SQL~\cite{li2024dawn} further improved robustness. More recent systems -- CHASE-SQL~\cite{pourreza2024chase}, Alpha-SQL~\cite{li2025alpha}, and Long Context NL2SQL~\cite{chung2025long} -- incorporate planning, candidate selection, and execution feedback for complex schemas.

Although existing methods perform strongly on Spider and BIRD, they struggle on Spider 2.0's larger schemas and more complex queries. Recent agentic approaches -- ReForCE~\cite{deng2025reforce} (self-refinement with majority-vote consensus), DSR-SQL~\cite{hao2025text} (dual-state reasoning), and AutoLink~\cite{wang2025autolink} (iterative schema linking) -- still achieve only 35.83\%, 35.28\%, and 34.92\% execution accuracy, respectively.

In contrast, \toolname\ introduces a novel approach that uses verifiable CTE-based agent views, generated from smaller schema chunks, as intermediate steps to guide the final SQL synthesis. This enables \toolname\ to achieves  a new state-of-the-art result of 70.38\% on Spider 2.0, while remaining competitive on Spider (85.59\%), BIRD (72.16\%) and KaggleDBQA (63.78\%).

\sstitle{Long Context Modeling for LLMs}
LLMs have finite context windows, making long inputs such as large documents and database schemas difficult to handle. Common strategies include extending context length (e.g.,from 2048 in GPT-3~\cite{brown2020language} to 128k in GPT-4~\cite{achiam2023gpt} and 400k tokens in GPT-5~\cite{singh2025openai}), though performance can still degrade due to context rot~\cite{hong2025context}; RAG, which retrieves relevant evidence~\cite{lewis2020retrieval,xu2023retrieval,wang2022text,lin2023train} but may omit necessary context~\cite{barnett2024seven}; and agentic frameworks such as CoA, where multiple agents process context segments and a manager aggregates their outputs~\cite{zhang2024chain}.

Inspired by CoA~\cite{zhang2024chain}, \toolname adapts long-context decomposition to Text-to-SQL by splitting large schemas into manageable \emph{chunks}, allowing agents to process table/column subsets sequentially rather than loading the full schema into a single prompt.
This makes schema filtering explicit, alleviates full-schema prompting, and improves robustness by validating intermediate agent views before composing the final SQL query.
\section{Problem Formulation}
\label{sec:model}

\subsection{Model and Problem Statement}

Given a natural language question \(\mathcal{Q}\), a database schema \(\mathcal{S}\), and optional external knowledge \(\mathcal{K}\), the goal of Text-to-SQL is to generate an executable SQL query \(\mathcal{Y}\) to answer the question \(\mathcal{Q}\). We define the database as
\(
\mathcal{D} = (\mathcal{T}, \mathcal{C}, \mathcal{R}),
\)
where \(\mathcal{T} = \{\mathcal{T}_1, \mathcal{T}_2, \ldots, \mathcal{T}_m\}\) is the set of \(m\) tables, and each table \(\mathcal{T}_i\) is associated with a column set
\(
\mathcal{C}_i = \{c^{i}_{1}, c^{i}_{2}, \ldots, c^{i}_{n_i}\},
\)
with \(n_i\) denoting the number of columns in \(\mathcal{T}_i\). Let \(\mathcal{C} = \{\mathcal{C}_1, \mathcal{C}_2, \ldots, \mathcal{C}_m\}\) be the collection of column sets for all tables, and let \(\mathcal{R}\) denote the foreign-key relations among tables. Then, the Text-to-SQL task can be formulated as
\begin{equation}
  \mathcal{Y}=f(\mathcal{Q}, \mathcal{S}, \mathcal{K} \mid \boldsymbol{\theta}),
\end{equation}
where the function $f(\cdot \mid \boldsymbol{\theta})$ can represent a model or pipeline system with the parameter $\boldsymbol{\theta}$, $\mathcal{Y}$ is SQL query can execute against the database instance $\mathcal{D}$ to get the data which use requires in $\mathcal{Q}$.
All important notations are summarized in ~\autoref{tab:notations}.

\begin{table}[htbp]
  \centering
  \vspace{-1em}
  \caption{Summary of Important Notations}
  \label{tab:notations}
  \vspace{-1em}
  \resizebox{0.46\textwidth}{!}{%
    \begin{tabular}{cl}
      \hline
      \textbf{Notation}                                                                                                                 & \textbf{Description}                                      \\ \hline
      $\mathcal{Q}$                                                                                                                     & Natural language question                                 \\
      $\mathcal{K}$                                                                                                                     & external knowledge evidence                               \\
      $\mathcal{D}$                                                                                                                     & A relational database                                     \\
      $\mathcal{S}$                                                                                                                     & Schema of the database $\mathcal{D}$                      \\
      $\mathcal{T}=\{\mathcal{T}_1, \mathcal{T}_2, \ldots, \mathcal{T}_m\}$                                                             & Set of tables in the database $\mathcal{D}$               \\
      $\mathcal{C}_i = \{c^{i}_{1}, c^{i}_{2}, \ldots, c^{i}_{n_i}\},$                                                                  & Set of columns of table $i$ in the database $\mathcal{D}$ \\
      $\Pi(\mathcal{S}) = \{\mathcal{S}^{1}_{\text{sub}}, \ldots, \mathcal{S}^{j}_{\text{sub}}, \ldots, \mathcal{S}^{k}_{\text{sub}}\}$ & Set of $k$ schema chunks                                  \\
      $ \mathcal{V}^{j}_{\text{CTE}}$                                                                                                   & CTE-based agent views for chunk $j$                       \\
      $\mathcal{J}^{j}$                                                                                                                 & JSON-based schema selection for chunk $j$                 \\
      \hline
    \end{tabular}
  }
  \vspace{-1em}
\end{table}

\subsection{Challenges}

Text-to-SQL systems in practical environments face several interrelated  challenges:

\begin{itemize}
  \item[(C1)] \textbf{Long-context overload.} Real-world databases may be accompanied by long external documents and large schemas with hundreds of tables and thousands of columns, making the input extremely long. To answer a single question, the system may need to process both the external knowledge and the full database schema, which can easily exceed the context window of LLMs. Even when all inputs fit, the long context can dilute attention, making it harder to identify relevant evidence and correctly ground schema elements. As a result, schema grounding weakens, increasing the risk of incomplete or incorrect SQL generation.

  \item[(C2)] \textbf{Limitations of schema reduction.} A common mitigation is to filter the schema by selecting tables and columns relevant to the question. However, in long-context settings, such filtering can reduce accuracy by removing useful schema elements, breaking valid join paths, or providing insufficient structure for correct SQL synthesis.

  \item[(C3)] \textbf{SQL complexity and error propagation.}
        Real-world questions often require complex SQL with multiple joins, nested conditions, aggregations, and multi-step reasoning.
        This complexity increases the likelihood of both schema-linking errors and semantic mistakes, and small errors in one part of the query can propagate to the final result.

  \item[(C4)] \textbf{Difficulty of debugging monolithic SQL.}
        When an LLM generates a single large SQL query, debugging is difficult: if execution fails or results are incorrect, it is often unclear which clause (e.g., join, filter, grouping) caused the issue.
        This makes iterative correction inefficient and unreliable.
\end{itemize}

\subsection{Framework Overview}

To address the challenges above, we propose \toolname, a multi-stage Text-to-SQL framework that uses executable CTE-based agent views as intermediate representations. Rather than generating a single monolithic SQL query over the full schema, \toolname constructs intermediate agent views that make schema filtering explicit, can be executed and validated against the database, and enable early error detection before final SQL synthesis. This design reduces full-schema prompting, improves error localization, and makes complex reasoning more tractable.

~\autoref{tab:method_comparison} compares \toolname with representative Text-to-SQL methods along three practical dimensions: single-candidate generation, zero-shot applicability, and support for large databases. Prompting-based methods such as DIN-SQL, DAIL-SQL, and MAC-SQL typically generate a single candidate, but are not designed for large-schema settings. Methods such as CHESS and MCS-SQL rely on selecting from multiple candidates, while recent Spider 2.0-oriented systems such as ReForCE and AutoLink handle large schemas but also depend on multi-candidate exploration. In contrast, \toolname supports single-candidate generation in a fully zero-shot setting while explicitly targeting large-database scenarios.  This combination is particularly appealing for practical deployments, where minimizing sampling/selection overhead is important, yet robust handling of long schemas and complex joins remains essential.

\begin{table}[t]
  \centering
  \small
  \caption{Comparison of \toolname with representative Text-to-SQL methods in terms of (i) single-candidate generation, (ii) zero-shot applicability, and (iii) ability to handle large databases (e.g., Spider 2.0 schemas).}
  \label{tab:method_comparison}
  \vspace{-.5em}
  \setlength{\tabcolsep}{3pt}
  \resizebox{0.48\textwidth}{!}{%
    \begin{tabular}{lccc}
      \toprule
      \textbf{Method}                                                  & \textbf{Single candidate} & \textbf{Zero-shot} & \textbf{Handle large DB} \\
      \midrule
      \makecell{DIN-SQL~\cite{pourreza2023dinsql}, DAIL-SQL~\cite{gao2023text} ,                                                                   \\ MAC-SQL~\cite{wang2024mac}} & \cmark & \xmark & \xmark \\
      \hline
      MCS-SQL~\cite{lee2025mcs}, CHESS~\cite{talaei2024chess}          & \xmark                    & \xmark             & \xmark                   \\
      Alpha-SQL~\cite{li2025alpha}                                     & \xmark                    & \cmark             & \xmark                   \\
      ReForCE~\cite{deng2025reforce}, AutoLink~\cite{wang2025autolink} & \xmark                    & \cmark             & \cmark                   \\
      \hline
      \toolname                                                        & \cmark                    & \cmark             & \cmark                   \\
      \bottomrule
    \end{tabular}
  }
  \vspace{-2.em}
\end{table}

\begin{figure*}[!h]
  \centering
  \includegraphics[width=0.99\linewidth]{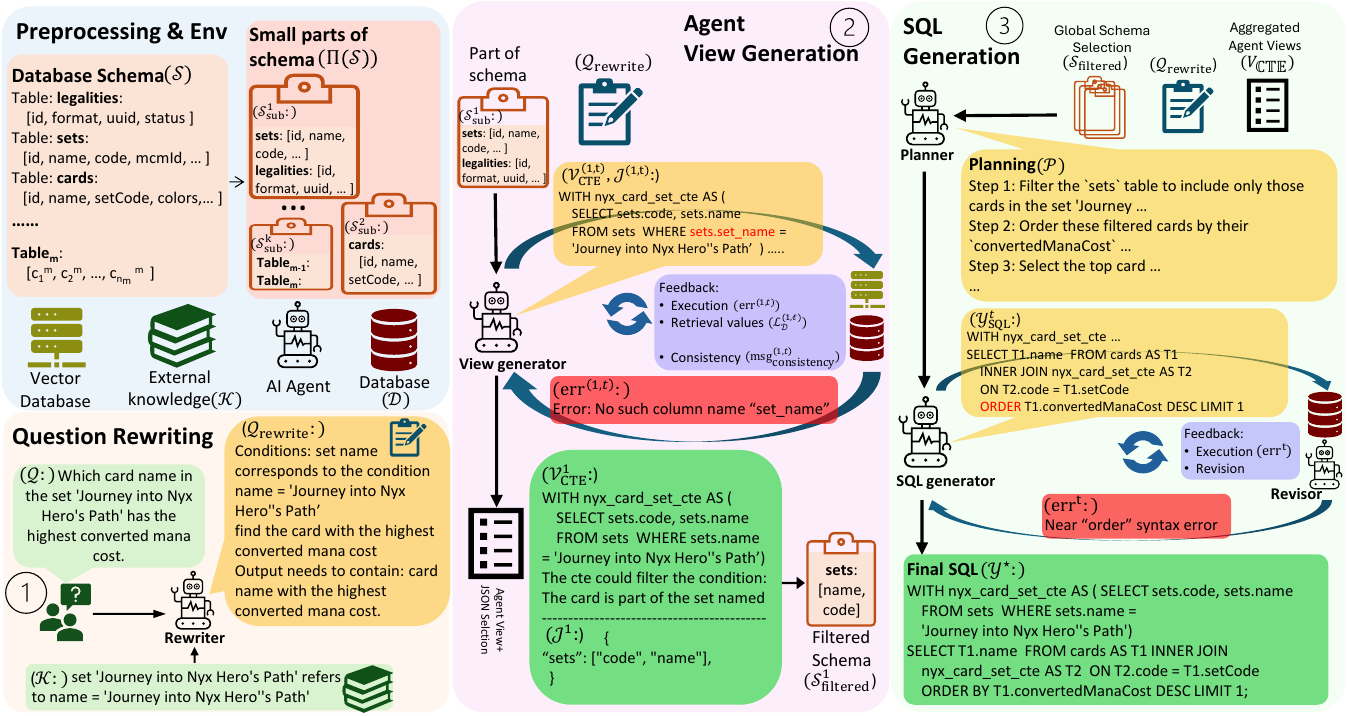}
  \vspace{-2pt}
  \caption{Overview of \toolname. The framework begins with preprocessing---schema splitting and vector database initialization for value preprocessing---followed by three main stages: (1) Question Rewriting
    which reformulates the input question, optionally using external knowledge, into a clearer and more explicit form; (2) Agent View Generation, which decomposes the long-context schema into smaller chunks and produces execution-validated intermediate CTE-based agent views that capture key reasoning steps while identifying relevant schema elements; and (3) SQL Generation, which uses three agents: a planner, a SQL generator, and a revision agent  to produce the final executable SQL query from the filtered schema and validated agent views.}
  \label{fig:framework}
  \vspace{-5pt}
\end{figure*}

In more detail, \toolname\ decomposes Text-to-SQL into three stages that progressively clarify the user intent, construct validated intermediate agent views from schema chunks, and synthesize the final executable SQL. As shown in~\autoref{fig:framework}, the framework consists of the following stages:
\begin{itemize}
  \item \textbf{Question Rewriting}: This stage rewrites the original question and external knowledge into a clearer and more explicit form. By extracting useful external knowledge, resolving underspecified references, and clarifying implicit constraints such as missing filters or aggregation criteria, it reduces linguistic ambiguity and improves downstream schema linking. The output is a clarified question that preserves the user's intent while making downstream reasoning more reliable, helping to mitigate challenges (C1) and (C3) by attention on relevant information to reduce the likelihood of semantic  errors in SQL generation.

  \item \textbf{Agent View Generation}: This stage uses multiple view-generation agents to process partial schema contexts, generating executable CTE-based intermediate views that incrementally capture the information needed to answer the clarified question. Each agent view is an executable Common Table Expressions (CTE) representing a specific reasoning step, such as identifying candidate entities, constructing join paths, applying filters, or computing aggregates. The agent views are validated through execution, or other database-level checks to ensure that the they are syntactically correct, structurally consistent, and semantically aligned with the intended query. They also provide useful insight into the database contents and the expected form of the final result. This stage addresses challenges (C1), (C2), (C3), and (C4) by decomposing long-context schemas into verifiable steps, making schema filtering explicit, and enabling early error detection before final SQL synthesis.
  \item \textbf{SQL Generation}: This stage generates the final SQL query using three agents: a \textit{planner}  that determines the query structure,  including selection target, filtering conditions, required \texttt{FROM} and \texttt{JOIN} clauses, and any nested structures; a \textit{SQL generator} that produces the query; and a \textit{revision} that checks the generated SQL for correctness and consistency with the question. Rather than using the full raw schema, these agents use only on the filtered schema elements and validated CTE-based agent views  from the previous stage, narrowing the search space and preventing hallucinated columns or incorrect joins. Building upon modular intermediate views makes the reasoning process transparent and verifiable, directly addressing challenges (C3) and (C4).
\end{itemize}

\section{Methodology}
\label{sec:method}

This section presents \toolname in detail, as illustrated in \autoref{fig:framework}. To address large-scale Text-to-SQL over real-world databases with hundreds of tables, our framework organizes the task into three stages: Question Rewriting, Agent View Generation, and SQL Generation. This decomposition reduces schema complexity and improves generation accuracy by dividing the long schema context into smaller and more manageable schema chunks. In particular, Question Rewriting clarifies the input question and extracts useful context, Agent View Generation builds validated intermediate agent views from partial schema contexts, and SQL Generation constructs the final SQL query using the filtered schema and validated CTE-based agent views.

\subsection{Preprocessing}
Before introducing the main stages of \toolname, we first apply a set of preprocessing steps to prepare the database.
In particular, the preprocessing procedure consists of three components: database schema compression, schema splitting into smaller chunks, and database value preprocessing.

\subsubsection{Database schema compression.}
\begin{figure}[ht]
  \centering
  \includegraphics[width=0.99\linewidth]{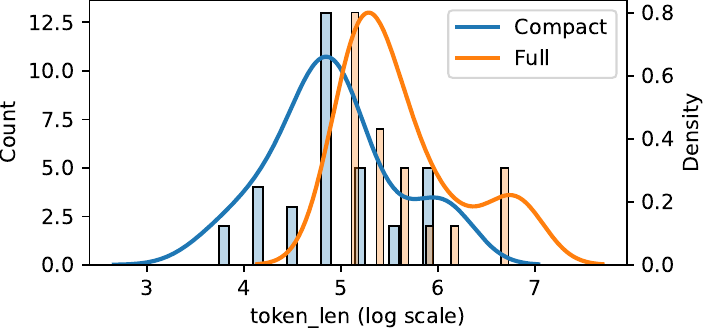}
  \caption{Token-length distribution of large database schemas before and after schema compression.}
  \label{fig:distribution_compress_token}
  \vspace{-1em}
\end{figure}

This step reduces the token length of Spider~2.0 schemas by grouping near-duplicate tables and columns into pattern-based clusters. Spider~2.0 often contains very large schemas with many similar tables and columns whose names differ only by numeric suffixes, such as year, month, or index. Directly serializing these schemas into prompts introduces substantial redundancy and may cause the model to miss important schema details.
To reduce this redundancy, we compress the schema by clustering redundant tables and columns based on shared naming patterns. Extending~\cite{deng2025reforce}, we further verify structural consistency before merging: tables are merged only when they share the same column sets and key information, and columns are grouped only when their types and metadata are consistent.
This process compresses large schemas by about 10$\times$ (see \autoref{fig:distribution_compress_token}), substantially reducing prompt length and improving efficiency.

\begin{example}
  In the \texttt{G360} database in Spider2-snow, the schema contains multiple tables such as \\
  \texttt{GOOGLE\_ANALYTICS\_SAMPLE.GA\_SESSIONS\_20170201}, \\
  \texttt{GOOGLE\_ANALYTICS\_SAMPLE.GA\_SESSIONS\_20170202},  ..., \\
  \texttt{GOOGLE\_ANALYTICS\_SAMPLE.GA\_SESSIONS\_20170228},  ... .\\  Since these tables have similar column sets and metadata, we group them into a single table cluster with the pattern \\ \texttt{GOOGLE\_ANALYTICS\_SAMPLE.GA\_SESSIONS\_\{NUM\}}.

  In the \texttt{WORLD\_BANK} database in Spider2-snow, the table\\ \texttt{WORLD\_BANK\_GLOBAL\_POPULATION.POPULATION\_BY\_COUNTRY}
  contains columns such as \texttt{year\_2003}, \texttt{year\_2004}, \texttt{year\_2005}, ...,\\ \texttt{year\_2022}. Because these columns share the same data type (\textit{Number}) and similar metadata, we group them into a single column cluster with the pattern  \texttt{year\_\{NUM\}}.'
\end{example}

\subsubsection{Database schema splitting}
\label{sec:split_databse_schema}
In large-scale benchmarks such as Spider~2.0, database schemas can be extremely long, reaching up to $\sim\!1.5$ million tokens, which exceeds the context window of current LLMs. To make schema grounding feasible for view generation stage of our framework, we partition the full schema into multiple smaller chunks, where each chunk contains one or more \emph{complete} table definitions.
Let the full database schema be $\mathcal{S}$. We partition it into $k$ chunks:
\begin{equation}
  \Pi(\mathcal{S}) = \{\mathcal{S}^{1}_{\text{sub}}, \mathcal{S}^{2}_{\text{sub}}, \ldots, \mathcal{S}^{k}_{\text{sub}}\}.
\end{equation}
Each chunk  $\mathcal{S}^{j}_{\text{sub}}$ contains a set of tables together with their complete column information and any foreign-key relations visible within that chunk :
\(
\mathcal{S}^{j}_{\text{sub}} =
\big(\mathcal{T}^{j},\, \mathcal{C}^{j},\, \mathcal{R}^{j}\big),
\)
where $\mathcal{T}^{j}$ denotes the set of tables, $\mathcal{C}^{j}$ the set of columns, and $\mathcal{R}^{j}$ the set of foreign-key relations visible in chunk $j$.

There are several possible strategies for splitting   $\mathcal{S}$: (i) assigning one table to each chunk, (ii) grouping tables by foreign-key connected components, or (iii) length-based partitioning. In practice, however, large schemas in Spider~2.0 do not provide  foreign-key information, so partitioning by connected components is not applicable.  We therefore use length-based partitioning, in which complete table definitions are packed into chunks such that the estimated token length of each chunk does not exceed a predefined limit \(B\):
\begin{equation}
  \forall \mathcal{S}^{j}_{\text{sub}} \in \Pi(\mathcal{S}), \quad \left|\mathcal{S}^{j}_{\text{sub}}\right| < B
\end{equation}
where $|\cdot|$ denotes the estimated token length of the corresponding schema chunk.

\vspace{-.5em}
\subsubsection{Database Value Preprocessing}
\label{subsubsec:database_value_preprocessing}
Accurately retrieving database values that match a user's natural-language intent is important for reliable Text-to-SQL generation~\cite{liu2025survey}. In practice, SQL queries typically reference only a small subset of values from potentially large databases, most often in filtering predicates such as \texttt{WHERE} clauses. A common challenge is the mismatch between how users mention an entity and how the corresponding value is stored in the database (e.g., a user writes ``USA'' while the database stores ``US''). Following~\cite{talaei2024chess, li2025alpha}, we adopt a two-stage pipeline with offline pre-processing and online retrieval to mitigate this discrepancy.

\sstitle{Offline Pre-processing}
In the offline stage, we index text-based entity values that are likely to vary in surface form, such as country names, organization names, abbreviations, and short categorical strings. We exclude primarily numeric values, such as phone numbers, dates, IDs, and postcodes. Each retained value is converted into a MinHash-based LSH signature~\cite{datar2004locality} and stored in a vector database for efficient retrieval.

\sstitle{Online Retrieval}
At query time, we extract candidate literals from the generated agent views. For each value, we compute its MinHash LSH signature~\cite{datar2004locality} and retrieve similar entries from the offline index. We then filter the candidates using edit-distance and semantic-similarity thresholds $(\tau_{edit}$, $\tau_{semantic})$, where semantic similarity is computed with  \textit{all-MiniLM-L6-v2}~\cite{wang2020minilm,reimers2019sentence}. The selected database values are then incorporated into the agent prompt, providing additional context for view generation and SQL synthesis.

\subsection{Question Rewriting}
\label{sec:rewrite_question}

This stage employs a single agent, $LLM_{Rewriter}$, to reformulate the user request into a clearer and more explicit form for downstream agents (e.g., the view generator, planner, and SQL generator). It takes as input the original natural-language question $\mathcal{Q}$ and optional external knowledge $\mathcal{K}$, and produces a rewritten question $\mathcal{Q}_{\text{rewrite}}$ that explicitly states the user intent, key constraints, and necessary domain assumptions.

External knowledge $\mathcal{K}$ provides schema-external information such as concept-to-column mappings, metric definitions (e.g., \\ \texttt{Average\_score}), business rules, and explanations of coded values. In benchmarks like BIRD, this knowledge is typically brief (often just 1--2 sentences). However, in large-scale datasets like Spider~2.0, $\mathcal{K}$ frequently consists of lengthy documents filled with noisy or question-irrelevant details. Providing the full $\mathcal{K}$ directly to downstream LLM agents increases prompt length and can distract the model, degrading Text-to-SQL reliability. Therefore, $LLM_{Rewriter}$ filters out irrelevant information from $\mathcal{K}$, integrates only the pertinent details, and rewrites the original question.
The rewritten question structures the key requirements explicitly, for example:
\[
  \mathcal{Q}_{\text{rewrite}}
  =
  \langle
  \text{task},
  \text{required outputs},
  \text{filters/constraints},
  \text{sorting/limit}
  \rangle.
\]

Formally, the overall rewriting objective is:
\begin{equation}
  \begin{aligned}
    \mathcal{Q}_{\text{rewrite}} & = LLM_{Rewriter}\!\big(\mathcal{Q}, \mathcal{K}\mid \mathcal{Q}\big), \\
    \text{s.t.}\quad             & \textsc{Intent}(\mathcal{Q}_{\text{rewrite}})
    =
    \textsc{Intent}(\mathcal{Q}, \mathcal{K}),
  \end{aligned}
  \label{eqn:rewrite_objective}
\end{equation}
Where $\textsc{Intent}(\cdot)$ is denoted as the expressed user intent in the question.
As a result, downstream agents receive only the clarified $\mathcal{Q}_{\text{rewrite}}$ rather than the raw question paired with the full $\mathcal{K}$, which reduces irrelevant context and improves robustness in subsequent reasoning and SQL generation.

\subsection{Agent View Generation}
\label{sec:cte_generation}

This stage is the core component for schema exploration and intermediate reasoning. It consists of multiple view-generator agents, denoted $LLM_{View}$, each receiving a partial schema chunk. Specifically, each agent takes as input (i) the rewritten question $\mathcal{Q}_{\text{rewrite}}$ from \autoref{sec:rewrite_question} and (ii) a schema chunk $\mathcal{S}^{j}_{\text{sub}}$ from the schema-splitting procedure in \autoref{sec:split_databse_schema}. Restricting each agent to a partial schema reduces prompt length and encourages the model to focus on a manageable subset of tables and columns, mitigating information overload in long-context settings.
Each view-generator agent, denoted $LLM_{View}^{j}$, produces two outputs:  (a) a CTE-based agent view $\mathcal{V}^{j}_{\text{CTE}}$ , consisting of one or more intermediate CTE queries with their natural-language rationale, and (b) a JSON schema selection  $\mathcal{J}^{j}$ identifying the candidate tables and columns needed to answer the question (see \autoref{fig:framework} for an example).
The generation process for each schema chunk $j$ is defined as:
\begin{equation}
  \mathcal{V}^{j}_{\text{CTE}},\, \mathcal{J}^{j}
  = LLM_{View}^{j}\!\left(\mathcal{Q}_{\text{rewrite}}, \mathcal{S}^{j}_{\text{sub}}\right),
\end{equation}
Each CTE query within $\mathcal{V}^{j}_{\text{CTE}}$ acts both as a data exploration step over the database and as a compressed logical representation of the reasoning behind table and column selection, encoding operations such as joins, filters, and aggregations in a modular form that can be directly reused during final SQL generation.
The JSON selection is defined as $\mathcal{J}^{j} = (\mathcal{T}_{\mathcal{J}}^{j}, \mathcal{C}_{\mathcal{J}}^{j})$, where  $\mathcal{T}_{\mathcal{J}}^{j}$ and $\mathcal{C}_{\mathcal{J}}^{j}$ denote the selected tables and columns, respectively.
These dual outputs serve two purposes: (i) it enables schema filtering for downstream SQL generation, and (ii) supporting automatic grounding quality checks by comparing the JSON selection against the schema elements actually used in the generated CTE.

To improve output quality, an iterative validation and repair procedure is applied to each agent view, covering keyword value extraction and retrieval, execution-based validation to fix syntax and schema-linking errors, and consistency checking between $\mathcal{V}^{j}_{\text{CTE}}$ and $\mathcal{J}^{j}$. Outputs are refined until valid or a maximum number of repair iterations is reached.
Once all schema chunks are processed, the validated agent views and JSON selections are aggregated into a set of executable agent views  and a global schema selection, passed to the downstream SQL Generation stage as compact, reusable intermediate reasoning context. The full procedure is described in \autoref{alg:cte-generation}, with each step detailed below.

\sstitle{Keyword Value Extraction and Retrieval}
In addition to producing intermediate queries, the CTE program exposes useful literals (e.g., string constants in \texttt{WHERE} clauses). Unlike prior work~\cite{talaei2024chess, li2025alpha} that extracts keywords directly via prompting, we use \texttt{sqlglot} to extract literals $\mathcal{L}^{j}_{\text{CTE}}$ from the generated CTE:
\begin{equation}
  \mathcal{L}^{j}_{\text{CTE}} = \textsc{Literal}\!\left(\mathcal{V}^{j}_{\text{CTE}}\right).
\end{equation}
For each literal $\ell \in \mathcal{L}^{j}_{\text{CTE}}$, we retrieve and normalize candidate database values denoted $\mathcal{L}^{(j,t)}_{\mathcal{D}}$ using the online value-matching procedure described in \autoref{subsubsec:database_value_preprocessing}.
The retrieved candidate set $\mathcal{L}^{(j,t)}_{\mathcal{D}}$ is then feedback to the CTE Agent to regenerate or refine $\mathcal{V}^{j}_{\text{cte}}$, improving value grounding and increasing the likelihood that the CTE-based agent view executes correctly.

\sstitle{Execution-based validation}
This step is to fix any syntax error in CTE query.
Each agent view contain one or more CTE queries,  which is executable and can be validated by running it against the database. We use \texttt{sqlglot} to parse the agent view and extract the underlying SQL for execution, which detects both syntax errors and schema-linking errors. If execution fails, the error message is fed back to the model , which is prompted to revise the agent view. We denote $\textsc{Valid}(\textsc{Exec}(\mathcal{V}^{j}_{\text{CTE}},\mathcal{D}))$ as the execution-based validation function, which returns \texttt{True} if $\mathcal{V}^{j}_{\text{CTE}}$ executes successfully on database instance $\mathcal{D}$ without error, and \texttt{False} otherwise. If there are any errors (denoted $\text{err}^{j}$), this error message will be feedback to the model to repair the CTE query. $\textsc{Valid}(\textsc{Exec}(\mathcal{V}^{j}_{\text{CTE}},\mathcal{D}))=\texttt{True}$ means that there is no error while executing the CTE query  $\text{err}^{j}=\varnothing$

\sstitle{Consistency checking}
We also enforce consistency between the CTE-based agent view and the JSON selection, denoted \\$\textsc{Consistent}(\mathcal{V}^{j}_{\text{CTE}}, \mathcal{J}^{j})$.
  Let $\mathcal{T}_\mathcal{V}^{j}$ and $\mathcal{C}_\mathcal{V}^{j}$ be the sets of tables and columns used by agent view CTE query $\mathcal{V}^{j}_{\text{CTE}}$,
  and JSON selection $\mathcal{J}^{j} = (\mathcal{T}_{\mathcal{J}}^{j}, \mathcal{C}_{\mathcal{J}}^{j})$. The consistency checking require:
  \begin{itemize}
    \item If the JSON output selects no columns and table, the agent must not generate any CTE-based view. It refer that this schema chunk is not relevant to answer the question, so there are not any table or column related to the question. \\
          \(
          \mathcal{T}_{\mathcal{J}}^{j} = \varnothing \land  \mathcal{C}_{\mathcal{J}}^{j} = \varnothing \;\Longrightarrow\; \mathcal{V}_{\mathrm{CTE}}^{j} = \varnothing.
          \)
    \item If a CTE-based view exists, the JSON output must include at least one selected column. It indicates that this provided schema chunk is related to the question, so JSON selection must choose at least one column or table which relevant to the question.
          \(
          \mathcal{V}_{\mathrm{CTE}}^{j} \neq \varnothing \;\Longrightarrow\; \mathcal{C}_{\mathcal{J}}^{j} \neq \varnothing \lor \mathcal{C}_{\mathcal{J}}^{j} = \varnothing
          \)
    \item All tables and columns referenced in the CTE-based view must appear in the JSON selection. This enforces that the CTE-based agent view uses only the tables and columns which are selected in JSON selection.  \\
          \(
          (\mathcal{T}_\mathcal{V}^{j} \subseteq \mathcal{T}_{\mathcal{J}}^{j})
          \land
          (\mathcal{C}_\mathcal{V}^{j} \subseteq \mathcal{C}_{\mathcal{J}}^{j})
          \)
  \end{itemize}

  After checking the consistency, the step will also feedback a consistency message denoted by $\texttt{msg}_\text{consistency}^{(j,t)}$ which could contain information about which table or column in CTE query is not selected in JSON selection, or any other inconsistency between CTE query and JSON selection. This message will be used to repair both CTE query and JSON selection in the next iteration.
  This consistency checking step ensure that the response of model is correct in logic. Since the model is required to express schema grounding in two forms--the CTE-based agent view and the JSON selection--any inconsistency between them may indicate a semantic error.
  Note that we allow redundancy: $\mathcal{J}^{j}$ may include additional tables/columns not used in the current CTE, since they can be useful when integrating evidence from other schema segments.

  \sstitle{Iterative validation and repair}
  The initial outputs $(\mathcal{V}^{(j,0)}_{\text{CTE}}$, $\mathcal{J}^{(j,0)})$ may contain errors, including SQL syntax errors, schema-linking errors such as invalid table or column names, and inconsistencies between the CTE and the JSON selection. We therefore apply an iterative validation-and-repair loop, regenerating both outputs until the CTE is executable and the selection is consistent:
  \begin{equation}
    \begin{aligned}
      \mathcal{V}^{(j,t+1)}_{\text{CTE}}, \mathcal{J}^{(j,t+1)}
      = LLM_{View}^{j}( & \mathcal{Q}_{\text{rewrite}}, \mathcal{S}^{j}_{\text{sub}}, \mathcal{V}^{(j,t)}_{\text{CTE}}, \mathcal{J}^{(j,t)}, \\
                        & \mathcal{L}^{j}_{\mathcal{D}}, \text{err}^{(j,t)},  \texttt{msg}_\text{consistency}^{(j,t)})
    \end{aligned}
  \end{equation}
  This loop continues until the validation conditions are satisfied or the maximum number of repair iterations  denoted $T_{\max}$ is reached.

  \sstitle{Aggregation}
  After all schema chunks have been processed, we aggregate the validated agent views and JSON selections:
  \begin{equation}
    \begin{aligned}
      \mathbb{V}_{\text{CTE}}
       & = \Big\{\mathcal{V}^{j}_{\text{CTE}} \;\Big|\;
      \textsc{Valid}\!\left(\textsc{Exec}(\mathcal{V}^{j}_{\text{CTE}}, \mathcal{D})\right)
      \land
      \textsc{Consistent}(\mathcal{V}^{j}_{\text{CTE}}, \mathcal{J}^{j})
      \Big\}_{j=1}^{k},                                 \\
      \mathbb{J}
       & = \Big\{\mathcal{J}^{j} \;\Big|\;
      \textsc{Consistent}(\mathcal{V}^{j}_{\text{CTE}}, \mathcal{J}^{j})
      \Big\}_{j=1}^{k}.
    \end{aligned}
  \end{equation}
  We then compute the globally selected tables and columns by taking the union over all chunks:
  \begin{equation}
    \mathcal{T}^{\star} = \bigcup_{j=1}^{k} \mathcal{T}_{\mathcal{J}}^{j},
    \qquad
    \mathcal{C}^{\star} = \bigcup_{j=1}^{k} \mathcal{C}_{\mathcal{J}}^{j}.
  \end{equation}
  Since the consistency check enforces
$(\mathcal{T}_\mathcal{V}^{j} \subseteq \mathcal{T}_{\mathcal{J}}^{j})$
  and
$(\mathcal{C}_\mathcal{V}^{j} \subseteq \mathcal{C}_{\mathcal{J}}^{j})$,
  the union of JSON selections forms a safe superset of all schema elements grounded by the validated CTEs.
  The global schema selection passed to the SQL Generation stage is:
  \(
  \mathcal{S}_{\text{filtered}} =
  \{\mathcal{T}^{\star},\, \mathcal{C}^{\star}\}.
  \)

  The following outputs  are passed to the downstream SQL Generation stage:
  \begin{itemize}
    \item \textbf{Aggregated Validated Agent Views} $\mathbb{V}_{\text{CTE}}$: validated, executable CTE-based agent view programs with their execution results and natural-language rationales explaining how each agent view contributes to the final answer.
    \item \textbf{Global Schema Selection} $\mathcal{S}_{\text{filtered}}$: a reduced schema derived from the aggregated JSON selections, containing only the relevant tables and columns.
  \end{itemize}

  Together, these outputs provide the SQL Generation stage with a compact and well-validated context, enabling it to reuse intermediate reasoning steps, avoid distraction from irrelevant schema elements, and leverage rationales to guide final query construction.

  \begin{algorithm}[ht]
    \caption{Agent view Generation Pipeline}
    \label{alg:cte-generation}
    \DontPrintSemicolon

    \KwIn{
    Rewritten question $\mathcal{Q}_{\text{rewrite}}$;\;
    Full schema $\mathcal{S}$;  and database instance $\mathcal{D}$;\;
    Token budget $B$ for schema segments;\;
    Max repair iterations $T_{\max}$;
    thresholds $\tau_{edit}$ and $\tau_{semantic}$
    }

    \KwOut{Aggregated Validated Agent Views $\mathbb{V}_{\text{CTE}}$; Global Schema Selection $\mathcal{S}_{\text{filtered}}$}
    \BlankLine
    $\{\mathcal{S}^{1}_{\text{sub}},\dots,\mathcal{S}^{k}_{\text{sub}}\} \leftarrow \textsc{SplitSchemaByTokenBudget}(\mathcal{S}, B)$\;
    $\mathbb{V}_{\text{CTE}} \leftarrow \varnothing$; $\mathbb{J} \leftarrow \varnothing$\;
    \For{$j \leftarrow 1$ \KwTo $k$}{
    \tcp{Initial generation}
    $(\mathcal{V}^{(j,0)}_{\text{CTE}}, \mathcal{J}^{(j,0)}) \leftarrow
      LLM_{View}^{j}(\mathcal{Q}_{\text{rewrite}}, \mathcal{S}^{j}_{\text{sub}})$\;

    $\text{response\_valid} \leftarrow \texttt{False}$\;
    $\mathcal{L}^{j}_{\mathcal{D}} \leftarrow \varnothing$ \tcp*{retrievel candidate values}
    $t \leftarrow 0$\;
    \tcp{Execution-based validation}
    \While{$t < T_{\max}$ \textbf{and} $\text{response\_valid} = \texttt{False}$}{
    \tcp{Keyword Value Extraction and Retrieval.}
    $\mathcal{L}^{(j,t)} \leftarrow \textsc{Literal}(\mathcal{V}^{(j,t)}_{\text{CTE}})$
    \ForEach{$\ell \in \mathcal{L}^{(j,t)}$}{
    $\mathcal{L}^{j}_{\mathcal{D}} \leftarrow \mathcal{L}^{j}_{\mathcal{D}} \cup  \textsc{Retrieve}(\ell, \tau_{edit}, \tau_{semantic})$\;
    }

    $(\mathbf{Z}^{(j,t)}, \text{err}^{(j,t)}) \leftarrow \textsc{Exec}(\mathcal{V}^{(j,t)}_{\text{CTE}}, \mathcal{D})$\;
    $\texttt{exec\_valid} \leftarrow (\text{err}^{(j,t)} = \varnothing)$\;

    \tcp{Consistency checking}
    $\texttt{is\_consistent}, \texttt{msg}_{\text{consistency}}^{(j,t)}\leftarrow~\textsc{Consistent}(\mathcal{V}^{(j,t)}_{\text{CTE}}, \mathcal{J}^{(j,t)})$\;
    \If{$\texttt{exec\_valid} \land \texttt{is\_consistent}$}{
      $\text{response\_valid} \leftarrow \texttt{True}$\;
    }
    \Else{
    $(\mathcal{V}^{(j,t+1)}_{\text{CTE}}, \mathcal{J}^{(j,t+1)}) \leftarrow
      LLM_{View}(\mathcal{Q}_{\text{rewrite}}, \mathcal{S}^{j}_{\text{sub}}, \mathcal{V}^{(j,t)}_{\text{CTE}}, \mathcal{J}^{(j,t)},$\;
    \hspace{1.5cm} $\mathcal{L}^{j}_{\mathcal{D}}, \text{err}^{(j,t)}, \texttt{msg}_\text{consistency}^{(j,t)})$
    }
    $t \leftarrow t + 1$\;
    }

    \If{$\text{response\_valid}=\texttt{True}$}{
    $\mathbb{V}_{\text{CTE}} \leftarrow \mathbb{V}_{\text{CTE}} \cup \{\mathcal{V}^{j}_{\text{CTE}}\}$ ;
    $\mathbb{J} \leftarrow \mathbb{J} \cup \{\mathcal{J}^{j}\}$\;
    }
    }
    \tcp{Aggregation}
    $\mathcal{T}^{\star} \leftarrow \bigcup_{\mathcal{T}_{\mathcal{J}}^{j}\in \mathbb{J}} \mathcal{T}_{\mathcal{J}}$ ;
    $\mathcal{C}^{\star} \leftarrow \bigcup_{\mathcal{C}_{\mathcal{J}}^{j}\in \mathbb{J}} \mathcal{C}_{\mathcal{J}}$\;

    $\mathcal{S}_{\text{filtered}} \leftarrow
      \{\mathcal{T}^{\star},\, \mathcal{C}^{\star}\}$\;

    \Return $(\mathbb{V}_{\text{CTE}}, \mathcal{S}_{\text{filtered}})$\;
  \end{algorithm}
  \vspace{-10pt}

  \subsection{SQL Generation}
  \label{sec:sql_generation}
  The \textbf{SQL Generation} stage synthesizes the final SQL query that answers the user question.
  It consists of three agents: a planner agent $LLM_{Planner}$, a SQL generator agent $LLM_{\text{SQL}}$, and a revision agent $LLM_{Revisor}$.
  This stage receives a compact, structured context from earlier stages: (i) the rewritten question $\mathcal{Q}_{\text{rewrite}}$ from \autoref{sec:rewrite_question}, (ii) the aggregated validated agent views $\mathbb{V}_{\text{CTE}}$ with their execution results and natural-language rationales from \autoref{sec:cte_generation},  and (iii) the global schema selection $\mathcal{S}_{\text{filtered}}$ containing only relevant tables and columns from \autoref{sec:cte_generation}.
  This reduced context keeps the agents focused on the necessary schema elements and lowers the risk of distraction from irrelevant database structures.
  The stage follows a three-step procedure: $LLM_{Planner}$ constructs a high-level query plan to reduce structural errors; $LLM_{\text{SQL}}$ generates the SQL query with execution feedback; and $LLM_{Revisor}$ verifies and revises the generated SQL for semantic correctness. The full procedure is described in \autoref{alg:sql-generation}, and each step is detailed below.

  \sstitle{Planning}
  Before writing SQL, $LLM_{Planner}$ constructs a high-level query plan to guide generation and reduce structural errors. The plan $\mathcal{P}$ specifies which CTEs and tables to use, required join paths, hidden formulas, necessary projections, and operations such as filtering, grouping, aggregation, and ordering:
  \begin{equation}
    \mathcal{P}
    =
    LLM_{Planner}\!\left(
    \mathcal{Q}_{\text{rewrite}},
    \mathbb{V}_{\text{CTE}},
    \mathcal{S}_{\text{filtered}}
    \right).
    \label{eqn:sql_agent_plan}
  \end{equation}

  \sstitle{Execution-based validation}
  Given the plan and the structured context from earlier stages, $LLM_{\text{SQL}}$ generates the final SQL query. Similar to the execution-based validation loop in \autoref{sec:cte_generation},  execution feedback is incorporated to iteratively repair syntax and schema-linking errors:
  \begin{equation}
    \mathcal{Y}_{\text{SQL}}^{t+1}
    =
    LLM_{\text{SQL}}\!\left(
    \mathcal{P},\mathcal{Q}_{\text{rewrite}},
    \mathbb{V}_{\text{CTE}},
    \mathcal{S}_{\text{filtered}},\mathcal{Y}_{\text{SQL}}^{t}, \text{err}^{(t)}
    \right)
    \label{eqn:sql_agent_generate_iter_feedback}
  \end{equation}
  where $\text{err}^{(t)}$ is the execution error message at iteration $t$. This loop terminates when the query $\mathcal{Y}_{\text{SQL}}$ executes successfully on the database instance $\mathcal{D}$ without error or when the maximum number of iterations $T_{\max}$ is reached.

  \sstitle{Semantic verification and revision}
  Successful execution does not guarantee semantic correctness. After obtaining an executable query from $LLM_{\text{SQL}}$, $LLM_{Revisor}$ is applied to verify whether the result matches the intent of $\mathcal{Q}_{\text{rewrite}}$:
  \begin{equation}
    \mathcal{Y}^{\star}_{\text{SQL}}
    =
    LLM_{Revisor}\!\left(
    \mathcal{Q}_{\text{rewrite}},
    \mathcal{Y}^{t}_{\text{SQL}}
    \right).
    \label{eqn:sql_agent_revise}
  \end{equation}
  If the revision agent confirms that $\mathcal{Y}_{\text{SQL}}$ is semantically correct, it acts as an identity mapping and returns the same query unchanged. Otherwise, $LLM_{Revisor}$ edits the query according to the corrective feedback and produces a revised query $\mathcal{Y}^{\star}_{\text{SQL}}$. This step reduces semantic errors that may survive execution-based validation.

  The SQL Generation stage outputs the final verified query $\mathcal{Y}^{\star}_{\text{SQL}}$ and its execution result as the system answer.

  \begin{algorithm}[ht]
    \caption{SQL Generation Pipeline}
    \label{alg:sql-generation}
    \DontPrintSemicolon
    \KwIn{Rewritten question $\mathcal{Q}_{\text{rewrite}}$;\;
    Aggregated Validated Agent Views $\mathbb{V}_{\text{CTE}}$;\;
    Global Schema Selection $\mathcal{S}_{\text{filtered}}$ and database $\mathcal{D}$;\;
    Max execution-repair iterations $T_{\max}$
    }
    \KwOut{Final SQL query $\mathcal{Y}^{\star}_{\text{SQL}}$ and execution result $\mathbf{Z}$}

    \BlankLine
    \tcp{Planning}
    $\mathcal{P} \leftarrow LLM_{\text{Planner}}\!\left(\mathcal{Q}_{\text{rewrite}}, \mathbb{V}_{\text{CTE}}, \mathcal{S}_{\text{filtered}}\right)$\;

    \BlankLine
    \tcp{Initial SQL generation from the plan}
    $\mathcal{Y}^{(0)}_{\text{SQL}} \leftarrow LLM_{\text{SQL}}(\mathcal{P}, \mathcal{Q}_{\text{rewrite}}, \mathbb{V}_{\text{CTE}}, \mathcal{S}_{\text{filtered}})$\;

    \BlankLine
    $t \leftarrow 0$\;
    \tcp{Execution-based validation}
    \While{$t < T_{\max}$}{
    $(\mathbf{Z}^{(t)}, \text{err}^{(t)}) \leftarrow \textsc{Exec}\!\left(\mathcal{Y}^{(t)}_{\text{SQL}}, \mathcal{D}\right)$\;

    \If{$\text{err}^{(t)} = \varnothing$}{
    $\mathcal{Y}^{\star}_{\text{SQL}}            \leftarrow \mathcal{Y}^{(t)}_{\text{SQL}} $\;
    \textbf{break}\;
    }

    $\mathcal{Y}^{(t+1)}_{\text{SQL}} \leftarrow~LLM_{\text{SQL}}\!\left(\mathcal{P}, \mathcal{Q}_{\text{rewrite}}, \mathbb{V}_{\text{CTE}}, \mathcal{S}_{\text{filtered}}, \mathcal{Y}^{(t)}_{\text{SQL}},\text{err}^{(t)}\right)$
    $t \leftarrow t + 1$\;
    }

    \tcp{Semantic verification and revision}

    $\mathcal{Y}^{\star}_{\text{SQL}} \leftarrow LLM_{\text{Revisor}}\!\left(\mathcal{Q}_{\text{rewrite}}, \mathcal{Y}^{t}_{\text{SQL}}\right)$\;
    $(\mathbf{Z}, \_) \leftarrow \textsc{Exec}\!\left(\mathcal{Y}^{\star}_{\text{SQL}}, \mathcal{D}\right)$\;

    \Return $(\mathcal{Y}^{\star}_{\text{SQL}}, \mathbf{Z})$
  \end{algorithm}
  \vspace{-1pt}

  \section{Experiments}
  \label{sec:exp}

  This section presents experiments evaluating the effectiveness, scalability, and reliability of \toolname. To systematically assess the contribution of each design choice, we organize our experiments around the following research questions.

  \begin{itemize}
    \item[(RQ1)] \textbf{Normal-scale performance:} How well does the framework perform on standard Text-to-SQL benchmarks: Spider, BIRD and KaggleDBQA?
    \item[(RQ2)] \textbf{Large-scale scalability:} How well does the framework perform on large scale, enterprise oriented benchmarks Spider~2.0-Snow?
    \item[(RQ3)] \textbf{Multi-Candidate SQL Generation:} What is the performance of \toolname under a multiple SQL candidates setting?
    \item[(RQ4)] \textbf{Ablation study} \textit{(Component contribution):} How do the key components--the rewriter, planner, revisor, and execution feedback for the view generator and SQL generator--contribute to overall performance?  \textit{(View Generation Output Contribution):} How do the two outputs of the Agent View Generation stage~\autoref{sec:cte_generation}, \textit{Aggregated Validated Agent Views} ($\mathbb{V}_{\text{CTE}}$) and \textit{Global Schema Selection} ($\mathcal{S}_{\text{filtered}}$) contribute to the correctness of the final SQL query?
    \item[(RQ5)] \textbf{Schema Filtering:} What is the accuracy of the Agent View Generation stage in identifying and filtering relevant tables and columns?
    \item[(RQ6)] \textbf{Component cost analysis}:What are the runtime and computational cost contributions of each agent in the proposed framework?
    \item[(RQ7)] \textbf{Error analysis}: What are the common error types and failure modes of the proposed framework?
    \item[(RQ8)]  \textbf{Hybrid LLM Configuration}:  What is the impact of using heterogeneous LLM backbones across agents, compared to a single shared backbone?
  \end{itemize}

  \subsection{Experimental settings}
  \label{sec:exp_settings}

  \sstitle{Databases and benchmarks}
  We evaluate our approach on both standard Text-to-SQL benchmarks and a large-scale, enterprise-oriented benchmark.
  \begin{itemize}
    \vspace{-1pt}
    \item Normal-scale databases.
          We evaluate on three widely used Text-to-SQL benchmarks: Spider~\cite{yu-etal-2018-spider}, which contains 200 multi-table databases across 138 diverse domains; BIRD~\cite{li2024can}, which includes 95 databases totaling 33.4\,GB from 37 professional domains; and KaggleDBQA~\cite{lee2021kaggledbqa}, a cross-domain benchmark of real-world web databases with domain-specific data types, original formatting, and unrestricted questions. For BIRD, we use the development version \textit{bird\_dev\_20251106}.

    \item Large-scale databases.
          To assess scalability on real-world schema, we evaluate on Spider2-Snow a Spider~2.0~\cite{lei2024spider} sub-task with 547 complex questions designed for enterprise-scale Text-to-SQL. Compared with standard academic benchmarks, it features much larger and more heterogeneous schemas, averaging about 800 columns per database, which makes schema understanding and SQL generation significantly more challenging. The Spider2-Snow requires generating SQL in the Snowflake dialect.
  \end{itemize}
  \sstitle{Evaluation metrics}
  \label{sec:exp_metrics}
  We evaluate model performance using \emph{Execution Accuracy} (EX). EX measures whether the SQL query predicted by the model produces the same execution result as the ground-truth SQL when run on the target database, and we use it as the primary indicator of correctness.
  It is important to note that the definition of EX differs across benchmarks. Spider, BIRD and KaggleDBQA use strict execution evaluation, requiring an exact match to the gold result in both content and column order. In contrast, Spider~2.0-Snow is more lenient, requiring only that the returned result contain the core information of the gold answer, without enforcing column order or penalizing extra columns.

  \sstitle{LLM configurations}
  Experiments are conducted using  several LLMs from closed-source flagship models (Gemini-3-Pro-Preview and GPT-5-Mini) and open-source models (Llama-3.3-70B and Qwen2.5-32B, short for Qwen2.5-Coder-32B). Unless otherwise specified, the decoding temperature is set to \(1.0\), with similarity thresholds \(\tau_{\text{edit}}=0.5\) and \(\tau_{\text{semantic}}=0.5\), and a maximum of \(T_{\max}=5\) execution feedback iterations. The token budget $B$ is set per model: \(10{,}000\) for Qwen2.5-32B, \(20{,}000\) for Llama-3.3-70B, \(80{,}000\) for GPT-5-Mini, and \(100{,}000\) for Gemini-3-Pro.

  \sstitle{Baselines}
  We compare our approach against representative Text-to-SQL systems, including both existing frameworks and pretrained models such as CodeS~\cite{li2024codes}, DIN-SQL~\cite{pourreza2023dinsql}, DAIL-SQL~\cite{gao2023text}, and CHESS~\cite{talaei2024chess}. For Spider~2.0-Snow, we additionally include baselines specifically designed for Spider~2.0~\cite{lei2024spider}, including Spider-Agent~\cite{lei2024spider}, AutoLink~\cite{wang2025autolink}, DSR-SQL~\cite{hao2025text} and ReFoRCE~\cite{deng2025reforce}.

  \subsection{Normal databases performance (RQ1)}
  To answer RQ1, \toolname is evaluated on three standard Text-to-SQL benchmarks: BIRD, Spider, and KaggleDBQA. \autoref{tab:spiderbird_results} reports Execution Accuracy (EX) on BIRD and Spider, comparing \toolname against both \emph{multi-candidate} and \emph{single-candidate} baselines, along with latency where available. \toolname with Gemini-3-Pro achieves the best BIRD result among single-candidate methods (\(72.16\%\)) and competitive Spider accuracy (\(85.59\%\)). With the same Qwen2.5-32B model, \toolname reaches \(67.99\%\) on BIRD, closely approaching Alpha-SQL-32B (\(69.70\%\))---a multi-candidate method with latency exceeding \(1000\)s---while avoiding expensive candidate search.
  \begin{table}[ht]
    \centering
    \small
    \caption{Execution Accuracy (EX) of our approach compared to baselines on the Spider and BIRD benchmarks.}
    \label{tab:spiderbird_results}
    \vspace{-1.5em}
    \setlength{\tabcolsep}{2pt}
    \resizebox{0.49\textwidth}{!}{%
      \begin{tabular}{llccc}
        \hline
        \textbf{Method}                        & \textbf{Infer model} & \textbf{BIRD}$\uparrow$ & \textbf{Spider}$\uparrow$ & \textbf{Latency(s)}$\downarrow$ \\
        \hline
        \multicolumn{3}{l}{\textit{Multi-candidate}}                                                                                                          \\
        MCS-SQL~\cite{lee2025mcs}              & GPT-4                & 63.36                   & 86.80                     & -                               \\
        Alpha-SQL-7B~\cite{li2025alpha}        & Qwen2.5-7B           & 66.80                   & 84.00                     & >1000                           \\
        Alpha-SQL-14B~\cite{li2025alpha}       & Qwen2.5-14B          & 68.70                   & 87.00                     & >1000                           \\
        Alpha-SQL-32B~\cite{li2025alpha}       & Qwen2.5-32B          & 69.70                   & -                         & >1000                           \\
        XiyanSQL-32B~\cite{gao2024xiyan}       & XiyanSQL-32B         & 73.34                   & -                         & -                               \\
        CHESS~\cite{talaei2024chess}           & GPT-4                & 68.31                   & 87.2                      & 118.61                          \\
        ZeroNL2SQL~\cite{fan2024combining}     & GPT-4                & -                       & 84.00                     & -                               \\
        \hline
        \multicolumn{3}{l}{\textit{Single-candidate}}                                                                                                         \\
        DIN-SQL~\cite{pourreza2023dinsql}      & GPT-4                & 50.72                   & 82.80                     & 24.09                           \\
        DAIL-SQL~\cite{gao2023text}            & GPT-4                & 55.90                   & 83.10                     & -                               \\
        MAC-SQL~\cite{wang2024mac}             & GPT-4                & 59.59                   & \textbf{86.75}            & 24.64                           \\
        SFT Llama2-13B~\cite{touvron2023llama} & Llama2-13B           & 53.91                   & 81.60                     & -                               \\
        CodeS-7B~\cite{li2024codes}            & CodeS-7B             & 57.17                   & 85.40                     & 1.87                            \\
        CodeS-15B~\cite{li2024codes}           & CodeS-15B            & 58.47                   & 84.90                     & 3.52                            \\
        XiyanSQL-32B~\cite{gao2024xiyan}       & XiyanSQL-32B         & 66.88                   & -                         & -                               \\
        XiyanSQL-7B~\cite{gao2024xiyan}        & XiyanSQL-7B          & 60.10                   & -                         & -                               \\
        LC-NL2SQL~\cite{chung2025long}         & Gemini-1.5-pro       & 67.41                   & 87.10                     & 12.3                            \\
        \hline
        \multicolumn{4}{l}{\textit{\toolname}} & 83.38                                                                                                        \\
        +                                      & Qwen2.5-32B          & 67.99                   & 84.33                     & --                              \\
        +                                      & Llama~3.3-70B        & 68.19                   & 83.66                     & --                              \\
        +                                      & Gemini-3-Pro         & \textbf{72.16}          & 85.59                     & --                              \\
        \hline
      \end{tabular}
    }%
    \vspace{-.5em}
  \end{table}
  \begin{table}[ht]
    \centering
    \caption{Execution Accuracy (EX) across different query hardness levels on the Spider benchmark.}
    \label{tab:spider_difficulty_splits}
    \vspace{-1em}
    \begin{tabular}{lcccc}
      \hline
      \textbf{Method}                          & \textbf{Easy}  & \textbf{Medium} & \textbf{Hard}  & \textbf{Extra} \\
      \hline
      CodeS-7B~\cite{li2024codes}              & 94.80          & \textbf{91.00}  & 75.30          & 66.90          \\
      CodeS-15B~\cite{li2024codes}             & \textbf{95.60} & 90.40           & 78.20          & 61.40          \\
      DIN-SQL(GPT-4)~\cite{pourreza2023dinsql} & 92.30          & 87.40           & 76.40          & 62.70          \\
      DAIL-SQL(GPT-4)~\cite{gao2023text}       & 91.50          & 90.10           & 75.30          & 62.70          \\
      Alpha-SQl-7B~\cite{li2025alpha}          & 94.00          & 89.20           & 76.40          & 63.30          \\
      Alpha-SQl-14B~\cite{li2025alpha}         & 94.00          & \textbf{91.00}  & 79.90          & \textbf{72.30} \\
      \hline
      \multicolumn{3}{l}{\textit{\toolname}}                                                                        \\
      +Qwen2.5-32B                             & 90.73          & 90.13           & 79.31          & 64.46          \\
      +Llama~3.3-70B                           & 91.13          & 85.43           & 82.76          & 68.67          \\
      +Gemini-3-Pro                            & 90.32          & 86.55           & \textbf{89.08} & 72.29          \\
      \hline
    \end{tabular}
  \end{table}
  \begin{table}[ht]
    \centering
    \caption{Execution Accuracy (EX) on BIRD dev set by difficulty level. For brevity, the abbreviation ”Mod” stands for ”Moderate” while ”Chall” denotes ”Challenging”}
    \label{tab:bird_difficulty_splits}
    \vspace{-1em}
    \setlength{\tabcolsep}{4pt}
    \begin{tabular}{l c c c}
      \hline
      \textbf{Method}                    & \textbf{Simple} & \textbf{Mod}   & \textbf{Chall} \\
      \hline
      DAIL-SQL(GPT-4)~\cite{gao2023text} & 63.00           & 45.60          & 43.10          \\
      MAC-SQL(GPT-4)~\cite{wang2024mac}  & 65.73           & 52.69          & 40.28          \\
      CodeS-7B~\cite{li2024codes}        & 64.60           & 46.90          & 40.30          \\
      CodeS-15B~\cite{li2024codes}       & 65.80           & 48.80          & 42.40          \\
      Alpha-SQL-7B~\cite{li2025alpha}    & 72.60           & 59.30          & 53.10          \\
      Alpha-SQL-14B~\cite{li2025alpha}   & 74.60           & 61.00          & 55.90          \\
      Alpha-SQL-32B~\cite{li2025alpha}   & 74.50           & 64.00          & 57.20          \\
      \hline
      \multicolumn{3}{l}{\textit{\toolname}}                                                 \\
      +Qwen2.5-32B                       & 72.37           & 64.44          & 52.41          \\
      +Llama~3.3-70B                     & 72.91           & 62.28          & 57.93          \\
      +Gemini-3-Pro                      & \textbf{76.60}  & \textbf{67.46} & \textbf{60.69} \\
      \hline
    \end{tabular}
  \end{table}

  \autoref{tab:spider_difficulty_splits} and \autoref{tab:bird_difficulty_splits} break down performance by difficulty level on Spider and BIRD, respectively. On Spider, \toolname performs competitively on Easy and Medium questions while achieving strong results on Hard and Extra queries. On BIRD, \toolname with Gemini-3-Pro attains the highest accuracy across all difficulty tiers.
  Notably, \toolname with open-source LLM (Qwen and Llama) outperforms both the fine-tuned CodeS model and the methods DAIL-SQL and MAC-SQL using closed-source GPT-4.

  \begin{table}[h!]
    \centering
    \caption{Execution Accuracy (EX) of our approach compared to baselines on the KaggleDBQA dataset.}
    \label{tab:kaggle_dbqa_results}
    \vspace{-1em}
    \resizebox{0.49\textwidth}{!}{%
      \begin{tabular}{l|ccccc}
        \hline
        \multirow{2}{*}{\textbf{Method}}                 & \multicolumn{5}{c}{\textbf{EX (\%)}}                                                                      \\
                                                         & \textbf{Easy}                        & \textbf{Medium} & \textbf{Hard}  & \textbf{Extra} & \textbf{All}   \\
        \hline
        DIN-SQL(gpt-4)~\cite{pourreza2023dinsql}         & -                                    & -               & -              & -              & 27.00          \\
        Incremental ICL~\cite{fan2024combining}(gpt-3.5) & -                                    & -               & -              & -              & 34.60          \\
        RESDSQL~\cite{li2023resdsql}                     & -                                    & -               & -              & -              & 31.90          \\
        RAT-SQL~\cite{wang-etal-2020-rat}                & -                                    & -               & -              & -              & 13.56          \\
        ZeroNL2SQL(gpt-4)~\cite{fan2024combining}        & -                                    & -               & -              & -              & 42.40          \\
        ZeroNL2SQL(gpt-3.5)~\cite{fan2024combining}      & -                                    & -               & -              & -              & 44.90          \\
        LC-NL2SQL~\cite{chung2025long}(Gemini-1.5-Pro)   & 65.94                                & 57.83           & 64.86          & 50.81          & 61.10          \\
        \hline
        \multicolumn{1}{l|}{\textit{\toolname}}                                                                                                                      \\
        +Llama~3.3-70B                                   & 65.96                                & 56.00           & 57.14          & 30.77          & 53.51          \\
        +Gemini-3-Pro                                    & \textbf{72.34}                       & \textbf{58.00}  & \textbf{67.35} & \textbf{56.41} & \textbf{63.78} \\
        \hline
      \end{tabular}
    }
  \end{table}
  \autoref{tab:kaggle_dbqa_results} evaluates \toolname on KaggleDBQA, a more challenging out-of-domain benchmark. \toolname with Gemini-3-Pro attains \(63.78\%\) overall EX, outperforming all listed baselines. Notably, \toolname with the open-source LLaMA model also surpasses most baselines that rely on closed-source LLMs, including ZeroNL2SQL with GPT-3.5, ZeroNL2SQL with GPT-4 and DIN-SQL with GPT-4.

  \subsection{Large scale database performance (RQ2)}
  To address RQ2 under more challenging conditions, we evaluate on Spider~2.0-Snow, which stress-tests text-to-SQL systems with substantially larger schemas and higher domain heterogeneity. \autoref{tab:spider2_snow_results} reports EX and average token usage on Spider~2.0-Snow. Standard methods such as CodeS-15B, DIN-SQL and LC-NL2SQL score \(0.00\%\), as they are not designed to handle large-scale Snowflake databases. The recent method ReFoRCE~\cite{deng2025reforce} reach only \(\sim35\%\). \toolname substantially outperforms all baselines, achieving up to \(70.38\%\) EX with Gemini-3-Pro, while consuming fewer tokens (\(92\text{k}\)) than DAIL-SQL~(\(124\text{k}\)) and ReFoRCE~(\(230\text{k}\)).
  This advantage holds consistently across different backbone models: using the same Qwen2.5-32B model, \toolname reaches \(25.05\%\) EX compared to \(5.48\%\) for Spider-Agent, demonstrating that \toolname consistently improves across different backbone LLMs.
  \begin{table}[!ht]
    \centering
    \small
    \caption{Execution Accuracy (EX) on Spider~2.0-Snow. Values in parentheses mark cases where the score reported in the published paper differs substantially from the official leaderboard; the parenthesized number corresponds to the leaderboard score.}
    \label{tab:spider2_snow_results}
    \vspace{-.75em}
    \setlength{\tabcolsep}{3pt}
    \resizebox{0.49\textwidth}{!}{%
      \begin{tabular}{l l r c}
        \hline
        \textbf{Method}                        & \textbf{Infer model} & \textbf{EX}$\uparrow$ & \textbf{Avg Tokens}$\downarrow$ \\
        \hline
        SFT CodeS-15B~\cite{li2024codes}       & CodeS-15B            & 0.00                  & --                              \\
        DIN-SQL~\cite{pourreza2023dinsql}      & GPT-4o               & 0.00                  & 32k                             \\
        CHESS~\cite{talaei2024chess}           & GPT-4o               & 1.28                  & --                              \\
        DAIL-SQL~\cite{gao2023text}            & GPT-4o               & 2.20                  & 124k                            \\
        LC-NL2SQL~\cite{chung2025long}         & gpt-5-mini           & 0.00                  & --                              \\
        Spider-Agent~\cite{lei2024spider}      & Qwen2.5-32B          & 5.48                  & --                              \\
        Spider-Agent~\cite{lei2024spider}      & GPT-4o               & 12.98                 & --                              \\
        Spider-Agent~\cite{lei2024spider}      & o3-mini              & 19.20                 & --                              \\
        Spider-Agent~\cite{lei2024spider}      & o1-preview           & 23.58                 & --                              \\
        ReFoRCE~\cite{deng2025reforce}         & o1-preview           & 31.26                 & $\approx$230k                   \\
        ReFoRCE~\cite{deng2025reforce}         & o3                   & 35.83 (62.89)         & $\approx$230k                   \\
        AutoLink~\cite{wang2025autolink}       & DeepSeek-R1          & 34.92 (54.84)         & --                              \\
        DSR-SQL~\cite{hao2025text}             & DeepSeek-R1          & 35.28 (63.80)         & --                              \\
        \hline
        \multicolumn{3}{l}{\textit{\toolname}} & 92k                                                                            \\
        +                                      & Qwen2.5-32B          & 25.05                 & --                              \\
        +                                      & Llama3.3-70B         & 27.42                 & --                              \\
        +                                      & gpt-5-mini           & 65.08                 & --                              \\
        +                                      & Gemini-3-Pro         & \textbf{70.38}        & --                              \\
        \hline
      \end{tabular}
    }
  \end{table}

  \begin{table}[!ht]
    \vspace{-.25em}
    \centering
    \small
    \caption{Execution Accuracy (EX) by difficulty level (easy/medium/hard) on Spider2-snow dataset}
    \label{tab:spider2_snow_difficulty_results}
    \vspace{-1em}
    \setlength{\tabcolsep}{3pt}
    \begin{tabular}{l l r r r}
      \hline
      \textbf{Method} & \textbf{Infer model} & \textbf{Easy}  & \textbf{Medium} & \textbf{Hard}  \\
      \hline
      \multicolumn{4}{l}{Spider-Agent~\cite{lei2024spider}}                                      \\
      +               & o1-preview           & 39.84          & 21.14           & 15.61          \\
      +               & o3-mini              & 31.25          & 18.29           & 11.56          \\
      +               & GPT-4o               & 24.22          & 11.38           & 6.94           \\
      +               & Qwen2.5-32B          & 4.47           & 2.31            & 5.48           \\
      \hline
      \multicolumn{4}{l}{\toolname}                                                              \\
      +               & Qwen2.5-32B          & 39.84          & 24.39           & 15.03          \\
      +               & Llama3.3-70B         & 38.28          & 32.11           & 12.71          \\
      +               & Gpt-5-mini           & \textbf{75.78} & 68.29           & 52.60          \\
      +               & Gemini-3-Pro         & 74.22          & \textbf{72.36}  & \textbf{64.74} \\
      \hline
    \end{tabular}
  \end{table}
  \autoref{tab:spider2_snow_difficulty_results} further breaks down performance by difficulty level. Spider-Agent (o1-preview) degrades sharply across easy, medium, and hard splits (\(39.84\%, 21.14\%, 15.61\%\)), whereas \toolname with Gemini-3-Pro maintains consistently high accuracy (\(74.22\%\), \(72.36\%\), \(64.74\%\)). The performance gap widens significantly on harder queries, demonstrating \toolname's stronger handling of compositional reasoning challenges in large-schema settings.

  \subsection{Multi-Candidate SQL Generation (RQ3)}
  \begin{figure}[ht]
    \centering
    \includegraphics[width=0.95\linewidth]{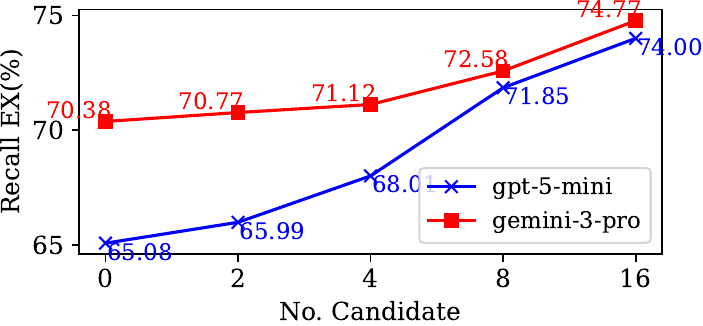}
    \vspace{-.75em}
    \caption{Recall Execution Accuracy on the Spider2.0-Snow benchmark under the multi-candidate setting.}
    \label{fig:multicandidate}
    \vspace{-1em}
  \end{figure}
  To answer RQ3, \toolname is evaluated in a multi-candidate setting on Spider2-Snow, where the SQL generation agent is prompted to output $k$ candidate queries. Performance is measured using \emph{recall} Execution Accuracy, which considers a question correctly answered if at least one of the $k$ generated candidates is executable correct. As shown in \autoref{fig:multicandidate}, increasing $k$ yields consistent gains for both backbones: Gemini-3-Pro improves from \(70.38\%\) at \(k{=}1\) to \(74.77\%\) at \(k{=}16\), while GPT-5-mini rises from \(65.08\%\) to \(74.00\%\) over the same range. These results indicate that generating more candidates increases the chance of producing at least one correct SQL query, improving robustness on challenging large-schema queries. This suggests that incorporating a candidate selector into the multi-candidate setting--one that reliably identifies the correct query among candidates--could further improve overall performance.
\begin{table*}[h]
  \centering
  \caption{Ablation study of major pipeline component--rewriter, planner, revisor, and execution feedback (exe feedback)--measured by Execution Accuracy (EX) across Spider2-Snow, Spider, BIRD, and KaggleDBQA. Llama-3.3 denotes Llama-3.3-70B.}
  \label{tab:ablation_components_all_datasets}
  \vspace{-1.em}
  \setlength{\tabcolsep}{3pt}
  \resizebox{0.99\textwidth}{!}{%
    \begin{tabular}{l | c c | c c | c c | cc}
      \hline
      \multirow{2}{*}{\textbf{Setting}}           &
      \multicolumn{2}{c|}{\textbf{Spider-2-snow}} &
      \multicolumn{2}{c|}{\textbf{Spider}}        &
      \multicolumn{2}{c|}{\textbf{BIRD}}          &
      \multicolumn{2}{c}{\textbf{KaggleDBQA}}                                                                                                                                             \\
                                                  &
      \makecell{GPT-5-mini}                       &
      \makecell{Gemini-3-Pro}                     &
      \makecell{Llama-3.3}                        &
      \makecell{Gemini-3-Pro}                     &
      \makecell{Llama-3.3}                        &
      \makecell{Gemini-3-Pro}                     &
      \makecell{Llama-3.3}                        &
      \makecell{Gemini-3-Pro}                                                                                                                                                             \\
      \hline
      w/o rewriter                                & 57.22          & 64.90          & 78.34          & 84.53          & 65.45          & 71.64          & 48.65          & 62.16          \\
      w/o exe feedback (SQL)                      & 50.82          & 63.99          & 80.33          & 84.91          & 64.54          & 71.77          & 51.35          & 61.62          \\
      w/o exe feedback (CTE+SQL)                  & 41.68          & 57.77          & 79.01          & 84.77          & 64.47          & 70.90          & 50.81          & 61.08          \\
      w/o planner                                 & 55.76          & 63.99          & 79.40          & 84.62          & 65.97          & 72.23          & 54.05          & 61.62          \\
      w/o reviseSQL                               & 60.15          & 65.08          & 82.50          & \textbf{87.14} & 67.47          & \textbf{72.82} & \textbf{54.59} & \textbf{65.41} \\
      \hline
      All components                              & \textbf{65.08} & \textbf{70.38} & \textbf{83.66} & 85.59          & \textbf{68.19} & 72.16          & 53.51          & 63.78          \\
      \hline
    \end{tabular}
  }
\end{table*}
\begin{table*}[h]
  \centering
  \small
  \caption{Ablation study on the contribution of each output from the Agent View Generation stage --  ($\mathbb{V}_{\text{CTE}}$) and  ($\mathcal{S}_{\text{filtered}}$) -- to SQL generation performance (EX) across Spider2-Snow, Spider, BIRD, and KaggleDBQA. Llama-3.3 denotes Llama-3.3-70B.}
  \label{tab:ablation_cte_info}
  \vspace{-1.1em}
  \setlength{\tabcolsep}{3pt}
  \resizebox{0.99\textwidth}{!}{%
    \begin{tabular}{l | cc | cc | cc | cc}
      \hline
      \multirow{2}{*}{\textbf{Information provided}}                            &
      \multicolumn{2}{c|}{\textbf{Spider2-snow}}                                &
      \multicolumn{2}{c|}{\textbf{Spider}}                                      &
      \multicolumn{2}{c|}{\textbf{BIRD}}                                        &
      \multicolumn{2}{c}{\textbf{KaggleDBQA}}                                                                                                                \\
                                                                                & \makecell{GPT-5-mini}   & \makecell{Gemini-3-Pro} &
      \makecell{Llama-3.3}                                                      & \makecell{Gemini-3-Pro} &
      \makecell{Llama-3.3}                                                      & \makecell{Gemini3-Pro}  & \makecell{Llama3.3}     & \makecell{Gemini3-Pro} \\
      \hline
      Aggregated Validated Agent Views ($\mathbb{V}_{\text{CTE}}$)              &
      58.68                                                                     & 64.72                   &
      79.83                                                                     & 85.42                   &
      67.41                                                                     & 71.64                   &
      51.89                                                                     & 62.70                                                                      \\
      Global Schema Selection ($\mathcal{S}_{\text{filtered}}$)                 &
      57.59                                                                     & 63.25                   &
      79.57                                                                     & 84.72                   &
      65.71                                                                     & 71.77                   &
      50.27                                                                     & 61.62                                                                      \\
      \hline
      All outputs ($\mathbb{V}_{\text{CTE}}$ + $\mathcal{S}_{\text{filtered}}$) &
      \textbf{65.08}                                                            & \textbf{70.38}          &
      \textbf{83.66}                                                            & \textbf{85.59}          &
      \textbf{68.19}                                                            & \textbf{72.16}          &
      \textbf{53.51}                                                            & \textbf{63.78}                                                             \\
      \hline
    \end{tabular}
  }
\end{table*}

\subsection{Ablation study (RQ4)}

\sstitle{Component contribution} To answer the first aspect of RQ4, a controlled ablation is conducted by disabling one component at a time---the rewriter, planner, revisor, and execution feedback for the view generator and SQL generator---while keeping all others intact.
\autoref{tab:ablation_components_all_datasets} shows that the full pipeline consistently outperforms all ablated variants across Spider2.0-Snow, Spider, BIRD, and KaggleDBQA. Removing execution feedback from both CTE and SQL generation causes the largest drops, with Gemini-3-Pro falling from \(70.38\%\) to \(57.77\%\) and GPT-5-mini from \(65.08\%\) to \(41.68\%\) on Spider2.0-Snow, confirming that multi-stage correction is critical for preventing error propagation. Notably, removing the revisor yields marginal improvements on Spider, BIRD, and KaggleDBQA (e.g., Gemini-3-Pro: \(85.59\%\rightarrow87.14\%\), \(72.16\%\rightarrow72.82\%\), \(63.78\%\rightarrow65.41\%\)).
the strict EX metric in these benchmarks requires an exact match of execution results in both content and column order (see \autoref{sec:exp_metrics}), meaning the revisor's edits---such as adding or removing columns---can cause a semantically correct query to be marked incorrect due to minor output differences.  Overall, these results confirm that each component contributes to final performance, with execution feedback playing the most critical role in large-schema settings.

\sstitle{View Generation Output
  Contribution}
To answer the second aspect of RQ4, this experiment isolates the contribution of each output from the View Generation stage to final SQL generation: (i) \emph{Aggregated Validated Agent Views} ($\mathbb{V}_{\text{CTE}}$) and (ii) \emph{Global Schema Selection} ($\mathcal{S}_{\text{filtered}}$)
As shown in \autoref{tab:ablation_cte_info}, combining both outputs consistently achieves the highest accuracy across all datasets and backbones. Consistent gains appear on Spider, BIRD, KaggleDBQA, and Spider2.0-Snow, confirming that schema selection and executable intermediate agent views are complementary and jointly necessary for accurate SQL generation.

\subsection{Schema Filtering (RQ5)}
To answer RQ5, this experiment evaluates schema filtering quality by measuring precision and recall of schema elements selected by each method. Precision measures whether selected elements are correct (few false positives), while recall measures whether all relevant elements are captured (few false negatives). Three baseline strategies are compared: (i) \textbf{MACSQL}~\cite{wang2024mac}, which prompts the model with the full schema to return relevant tables and columns; (ii) \textbf{CHESS}~\cite{talaei2024chess}, which evaluates table relevance one at a time before selecting columns from relevant tables; and (iii) an \textbf{encoder--decoder} schema selector~\cite{li2024codes}, which predicts relevant schema items from the question and schema as input.
\begin{table}[!ht]
  \vspace{-.1em}
  \centering
  \small
  \caption{Precision and recall of schema filtering across methods and inference models in Spider2-snow dataset.}
  \label{tab:schema_filtering}
  \vspace{-1.45em}
  \setlength{\tabcolsep}{3pt}
  \begin{tabular}{l l c c}
    \hline
    \textbf{Method}                                        & \textbf{Infer model} & \textbf{Precision} & \textbf{Recall} \\
    \hline
    \multicolumn{4}{l}{MACSQL~\cite{wang2024mac}}                                                                        \\
    +                                                      & Qwen2.5              & 62.00              & 63.88           \\
    +                                                      & Llama~3.3            & 68.30              & 68.91           \\
    +                                                      & GPT-5-Mini           & 59.41              & 79.22           \\
    +                                                      & Gemini-3-Pro         & 70.62              & 86.45           \\
    \multicolumn{4}{l}{CHESS~\cite{talaei2024chess}}                                                                     \\
    +                                                      & Qwen2.5              & 49.03              & 77.78           \\
    +                                                      & Llama~3.3            & 27.99              & 76.70           \\
    +                                                      & GPT-5-Mini           & 27.50              & \textbf{91.12}  \\
    +                                                      & Gemini-3-Pro         & 36.87              & 90.58           \\
    \multicolumn{2}{l}{Encoder-Decoder~\cite{li2024codes}} & 41.50                & 79.60                                \\
    \hline
    \multicolumn{4}{l}{CTE generation in \toolname}                                                                      \\
    +                                                      & Qwen2.5              & 69.74              & 69.27           \\
    +                                                      & Llama~3.3            & 77.76              & 74.67           \\
    +                                                      & GPT-5-Mini           & 81.41              & 82.70           \\
    +                                                      & Gemini-3-Pro         & \textbf{86.34}     & 81.59           \\
    \hline
  \end{tabular}
\end{table}
As shown in~\autoref{tab:schema_filtering}, CHESS achieves high recall (\(90.58\text{--}91.12\%\)) but low precision (\(27.50\text{--}36.87\%\)), indicating excessive false positives. MACSQL achieves a better balance (\(59.41\text{--}70.62\%\) precision, \(63.88\text{--}86.45\%\) recall), while the encoder--decoder baseline yields \(41.50\%\) precision and \(79.60\%\) recall. \toolname's CTE generation achieves the highest precision across all backbones (\(81.41\text{--}86.34\%\) with proprietary models; \(69.74\text{--}77.76\%\) with open-weight models) while maintaining competitive recall (\(81.59\text{--}82.70\%\)), confirming that CTE-based intermediate reasoning produces cleaner and more reliable schema filtering than prompting or learned approaches.

\subsection{Component cost analysis (RQ6)}
\label{sec:cost_analysis}
\begin{figure}[ht]
  \centering
  \includegraphics[width=0.99\linewidth]{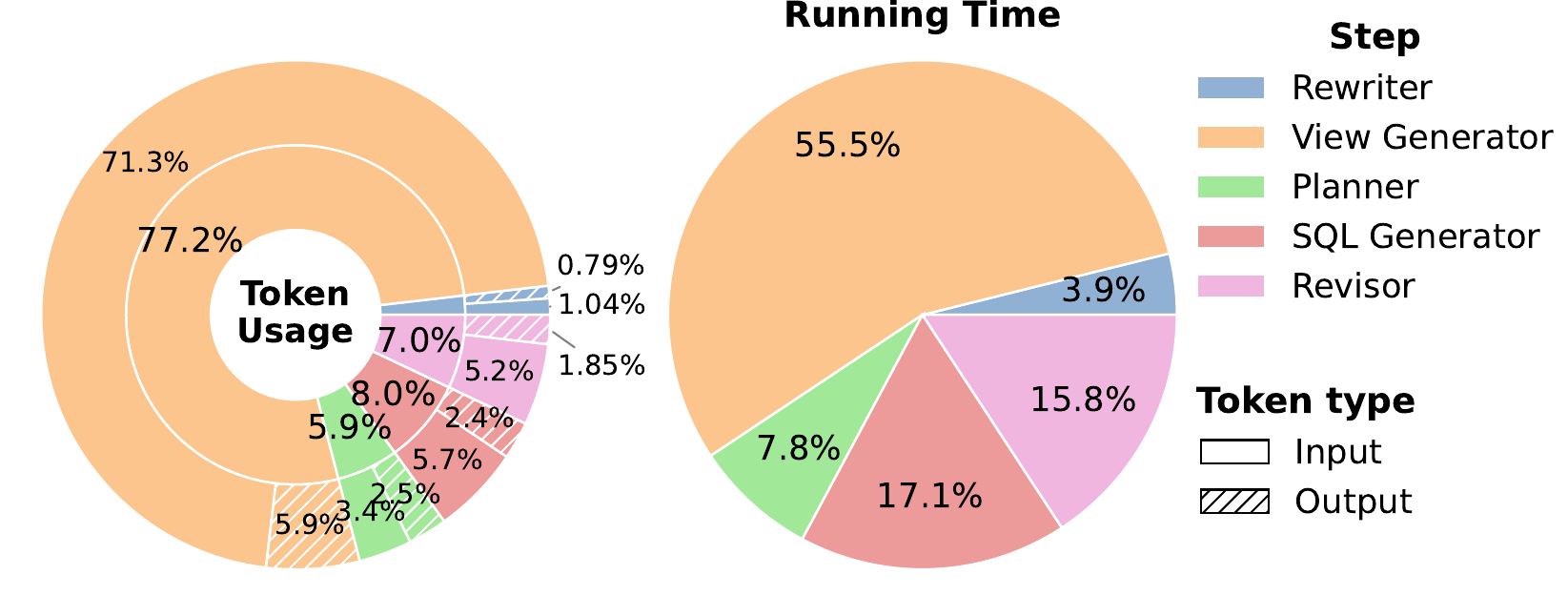}
  \vspace{-1em}
  \caption{Token usage and runtime  breakdown by pipeline agent for \toolname with Gemini-3-Pro on Spider2.0-snow. The left nested donut chart shows the proportion of token usage, while the right pie chart shows the proportion of total runtime spent on each agent.}
  \label{fig:token_time_breakdown}
  \vspace{-.75em}
\end{figure}

To answer RQ6, \autoref{fig:token_time_breakdown} presents the token usage and runtime breakdown per pipeline agent for \toolname with Gemini-3-Pro on Spider2.0-Snow.
The view generator agent dominates both dimensions, consuming \(77.2\%\) of total tokens (\(71.3\%\) input, \(5.9\%\) output) and \(55.5\%\) of total runtime, reflecting the cost of processing multiple schema chunks with iterative validation. The SQL generator and revisor agents account for \(8.4\%\) and \(7.0\%\) of token usage, and \(17.1\%\) and \(15.8\%\) of runtime, respectively. The planner contributes \(5.9\%\) of tokens and \(7.8\%\) of runtime, while the rewriter is the lightest stage at \(1.83\%\) of tokens and \(3.9\%\) of runtime. Overall, the cost profile is heavily concentrated in the view  generation agent, which is expected given its role as the core  intermediate reasoning, schema exploration and schema filtering component, while the remaining agents impose relatively modest overhead.

\subsection{Error Analysis (RQ7)}
To answer RQ7, a human evaluation is conducted on the failure cases of \toolname with Gemini-3-Pro on Spider2-Snow.

\begin{figure}[ht]
  \vspace{-1em}
  \centering
  \includegraphics[width=0.5\linewidth]{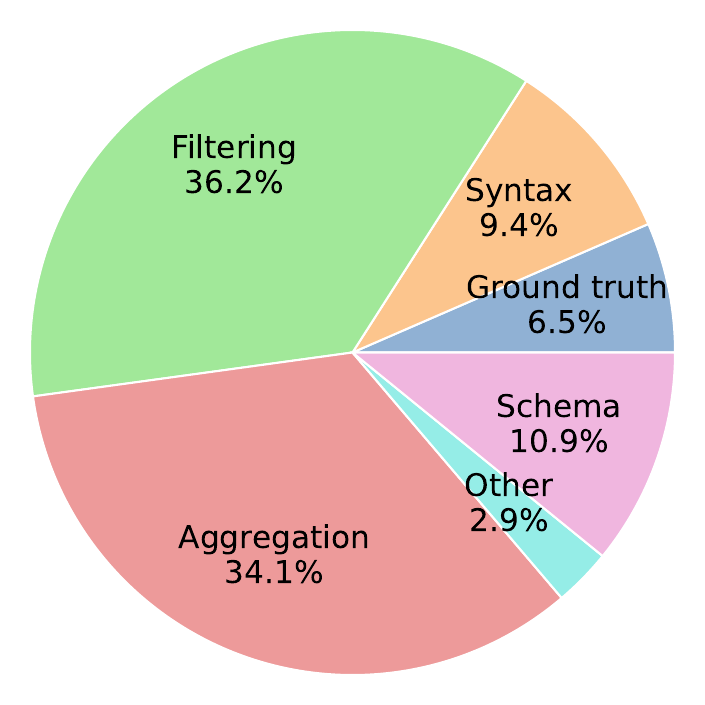}
  \vspace{-1em}
  \caption{Error taxonomy of \toolname with Gemini-3-Pro on Spider2-snow.}
  \label{fig:error_taxonomy}
  \vspace{-1em}
\end{figure}

\autoref{fig:error_taxonomy} shows the resulting error taxonomy. Filtering (\(36.2\%\)) and aggregation (\(34.1\%\)) errors dominate, together accounting for over \(70\%\) of failures, indicating that complex filter conditions and aggregation logic over large schemas remain the primary challenges. Schema errors contribute \(10.9\%\), while syntax errors account for only \(9.4\%\), confirming that execution feedback effectively suppresses most malformed queries. The remaining errors correspond to lacking ground truth  (\(6.5\%\)) and miscellaneous cases (\(2.9\%\)). These results suggest that future work should prioritize more precise filter and aggregation reasoning.

\subsection{Hybrid LLM configuration (RQ8)}
To answer RQ8, \autoref{tab:hybrid_llm} evaluates the effect of assigning different LLM backbones to different agents, compared to a single unified backbone.
\begin{table}[ht]
  \centering
  \small
  \caption{Execution Accuracy (EX) on Spider2-Snow under hybrid LLM configurations, where different backbone models are assigned to different agents.}
  \label{tab:hybrid_llm}
  \vspace{-1em}
  \setlength{\tabcolsep}{3pt}
  \resizebox{0.49\textwidth}{!}{%
    \begin{tabular}{c c c c c c}
      \hline
      \multirow{2}{*}{\textbf{Rewriter}} & \multirow{2}{*}{\textbf{View Gen.}} & \multicolumn{3}{c}{\textbf{SQL Generation}} & \multirow{2}{*}{\textbf{EX}$\uparrow$}                                     \\
                                         &                                     & \textbf{Planner}                            & \textbf{SQL Gen.}                      & \textbf{Revisor} &                \\
      \hline
      \multicolumn{6}{l}{\textit{Homogeneous}}                                                                                                                                                            \\
      Llama-3.3                          & Llama-3.3                           & Llama-3.3                                   & Llama-3.3                              & Llama-3.3        & 27.42          \\
      GPT-5-mini                         & GPT-5-mini                          & GPT-5-mini                                  & GPT-5-mini                             & GPT-5-mini       & 65.08          \\
      Gemini-3-Pro                       & Gemini-3-Pro                        & Gemini-3-Pro                                & Gemini-3-Pro                           & Gemini-3-Pro     & \textbf{70.38} \\
      \hline
      \multicolumn{6}{l}{\textit{Hybrid}}                                                                                                                                                                 \\
      Llama-3.3                          & Llama-3.3                           & Llama-3.3                                   & GPT-5-mini                             & GPT-5-mini       & 40.40          \\
      GPT-5-mini                         & Llama-3.3                           & GPT-5-mini                                  & GPT-5-mini                             & GPT-5-mini       & 42.96          \\
      Gemini-3-Pro                       & Llama-3.3                           & Gemini-3-Pro                                & Gemini-3-Pro                           & Gemini-3-Pro     & 55.21          \\
      Gemini-3-Pro                       & GPT-5-mini                          & Gemini-3-Pro                                & Gemini-3-Pro                           & Gemini-3-Pro     & 61.60          \\
      \hline
    \end{tabular}
  }
\end{table}
Homogeneous configurations with strong LLM backbones achieve the best results,  all-GPT-5-mini \(65.08\%\), and all-Gemini-3-Pro \(70.38\%\) on Spider2-Snow. Replacing only the SQL generation stage with a stronger model while retaining Llama-3.3 for the rewriter and CTE generator improves performance from \(27.42\%\) to \(40.40\%\), demonstrating that a stronger SQL generator can better exploit the filtered context produced by the CTE stage. This is also cost-efficient, since the CTE generation stage dominates token usage~\autoref{sec:cost_analysis}, meaning the expensive backbone is only applied where it matters most. However, mixing backbones does not consistently improve performance: using Gemini-3-Pro for the rewriter and CTE generator but GPT-5-mini for view generation yields only \(61.60\%\), falling below the homogeneous GPT-5-mini baseline (\(65.08\%\)), suggesting that misalignment between the CTE generator and SQL generator can degrade performance even when individual components are strong.
\section{Conclusions}
\label{sec:conclusion}

This paper presents \toolname, a multi-agent Text-to-SQL framework designed for robust SQL generation across both standard and large-scale database settings. \toolname decomposes the generation process into three stages: (1) \textit{Question Rewriting} rewrites the input question (and optional external knowledge) into a clearer and more concise form; (2) \textit{Agent View Generation} processes schema chunks, producing validated CTE-based agent views and a global schema selection; and (3) \textit{SQL Generation} composes the validated agent views and filtered schema to generate the final executable SQL query.
Extensive experiments demonstrate that \toolname achieves competitive or state-of-the-art execution accuracy across Spider, BIRD, KaggleDBQA, and Spider2.0-Snow. Notably, on Spider2.0-Snow, where large schemas pose significant challenges to conventional pipelines, \toolname achieves \(70.38\%\) EX with Gemini-3-Pro, surpassing the previous methods and proving that processing schema chunks into verifiable intermediate views provides a highly scalable solution.
Future work will focus on enhancing reasoning mechanisms for complex filtering and aggregation, which remain our primary sources of error. Additionally, we aim to integrate robust candidate selection and optimize heterogeneous LLM alignments to further improve performance and cost-efficiency.

\balance

%%% -*-BibTeX-*-
%%% Do NOT edit. File created by BibTeX with style
%%% ACM-Reference-Format-Journals [18-Jan-2012].


\begin{thebibliography}{48}

%%% ====================================================================
%%% NOTE TO THE USER: you can override these defaults by providing
%%% customized versions of any of these macros before the \bibliography
%%% command.  Each of them MUST provide its own final punctuation,
%%% except for \shownote{}, \showDOI{}, and \showURL{}.  The latter two
%%% do not use final punctuation, in order to avoid confusing it with
%%% the Web address.
%%%
%%% To suppress output of a particular field, define its macro to expand
%%% to an empty string, or better, \unskip, like this:
%%%
%%% \newcommand{\showDOI}[1]{\unskip}   % LaTeX syntax
%%%
%%% \def \showDOI #1{\unskip}           % plain TeX syntax
%%%
%%% ====================================================================

\ifx \showCODEN    \undefined \def \showCODEN     #1{\unskip}     \fi
\ifx \showDOI      \undefined \def \showDOI       #1{#1}\fi
\ifx \showISBNx    \undefined \def \showISBNx     #1{\unskip}     \fi
\ifx \showISBNxiii \undefined \def \showISBNxiii  #1{\unskip}     \fi
\ifx \showISSN     \undefined \def \showISSN      #1{\unskip}     \fi
\ifx \showLCCN     \undefined \def \showLCCN      #1{\unskip}     \fi
\ifx \shownote     \undefined \def \shownote      #1{#1}          \fi
\ifx \showarticletitle \undefined \def \showarticletitle #1{#1}   \fi
\ifx \showURL      \undefined \def \showURL       {\relax}        \fi
% The following commands are used for tagged output and should be
% invisible to TeX
\providecommand\bibfield[2]{#2}
\providecommand\bibinfo[2]{#2}
\providecommand\natexlab[1]{#1}
\providecommand\showeprint[2][]{arXiv:#2}

\bibitem[\protect\citeauthoryear{Achiam, Adler, Agarwal, Ahmad, Akkaya, Aleman,
  Almeida, Altenschmidt, Altman, Anadkat, et~al\mbox{.}}{Achiam
  et~al\mbox{.}}{2023}]%
        {achiam2023gpt}
\bibfield{author}{\bibinfo{person}{Josh Achiam}, \bibinfo{person}{Steven
  Adler}, \bibinfo{person}{Sandhini Agarwal}, \bibinfo{person}{Lama Ahmad},
  \bibinfo{person}{Ilge Akkaya}, \bibinfo{person}{Florencia~Leoni Aleman},
  \bibinfo{person}{Diogo Almeida}, \bibinfo{person}{Janko Altenschmidt},
  \bibinfo{person}{Sam Altman}, \bibinfo{person}{Shyamal Anadkat},
  {et~al\mbox{.}}} \bibinfo{year}{2023}\natexlab{}.
\newblock \showarticletitle{Gpt-4 technical report}.
\newblock \bibinfo{journal}{\emph{arXiv preprint arXiv:2303.08774}}
  (\bibinfo{year}{2023}).
\newblock


\bibitem[\protect\citeauthoryear{Barnett, Kurniawan, Thudumu, Brannelly, and
  Abdelrazek}{Barnett et~al\mbox{.}}{2024}]%
        {barnett2024seven}
\bibfield{author}{\bibinfo{person}{Scott Barnett}, \bibinfo{person}{Stefanus
  Kurniawan}, \bibinfo{person}{Srikanth Thudumu}, \bibinfo{person}{Zach
  Brannelly}, {and} \bibinfo{person}{Mohamed Abdelrazek}.}
  \bibinfo{year}{2024}\natexlab{}.
\newblock \showarticletitle{Seven failure points when engineering a retrieval
  augmented generation system}. In \bibinfo{booktitle}{\emph{CAIN}}.
  \bibinfo{pages}{194--199}.
\newblock


\bibitem[\protect\citeauthoryear{Brown, Mann, Ryder, Subbiah, Kaplan, Dhariwal,
  Neelakantan, Shyam, Sastry, Askell, et~al\mbox{.}}{Brown
  et~al\mbox{.}}{2020}]%
        {brown2020language}
\bibfield{author}{\bibinfo{person}{Tom Brown}, \bibinfo{person}{Benjamin Mann},
  \bibinfo{person}{Nick Ryder}, \bibinfo{person}{Melanie Subbiah},
  \bibinfo{person}{Jared~D Kaplan}, \bibinfo{person}{Prafulla Dhariwal},
  \bibinfo{person}{Arvind Neelakantan}, \bibinfo{person}{Pranav Shyam},
  \bibinfo{person}{Girish Sastry}, \bibinfo{person}{Amanda Askell},
  {et~al\mbox{.}}} \bibinfo{year}{2020}\natexlab{}.
\newblock \showarticletitle{Language models are few-shot learners}.
\newblock \bibinfo{journal}{\emph{NeurIPS}}  \bibinfo{volume}{33}
  (\bibinfo{year}{2020}), \bibinfo{pages}{1877--1901}.
\newblock


\bibitem[\protect\citeauthoryear{Cao, Chen, Chen, Zhao, Zhu, and Yu}{Cao
  et~al\mbox{.}}{2021}]%
        {cao-etal-2021-lgesql}
\bibfield{author}{\bibinfo{person}{Ruisheng Cao}, \bibinfo{person}{Lu Chen},
  \bibinfo{person}{Zhi Chen}, \bibinfo{person}{Yanbin Zhao},
  \bibinfo{person}{Su Zhu}, {and} \bibinfo{person}{Kai Yu}.}
  \bibinfo{year}{2021}\natexlab{}.
\newblock \showarticletitle{{LGESQL}: Line Graph Enhanced Text-to-{SQL} Model
  with Mixed Local and Non-Local Relations}. In
  \bibinfo{booktitle}{\emph{IJCNLP}}. \bibinfo{pages}{2541--2555}.
\newblock


\bibitem[\protect\citeauthoryear{Chung, Kakkar, Gan, Milne, and Ozcan}{Chung
  et~al\mbox{.}}{2025}]%
        {chung2025long}
\bibfield{author}{\bibinfo{person}{Yeounoh Chung}, \bibinfo{person}{Gaurav~T
  Kakkar}, \bibinfo{person}{Yu Gan}, \bibinfo{person}{Brenton Milne}, {and}
  \bibinfo{person}{Fatma Ozcan}.} \bibinfo{year}{2025}\natexlab{}.
\newblock \showarticletitle{Is long context all you need? leveraging LLM's
  extended context for NL2SQL}.
\newblock \bibinfo{journal}{\emph{PVLDB}} \bibinfo{volume}{18},
  \bibinfo{number}{8} (\bibinfo{year}{2025}), \bibinfo{pages}{2735--2747}.
\newblock


\bibitem[\protect\citeauthoryear{Datar, Immorlica, Indyk, and Mirrokni}{Datar
  et~al\mbox{.}}{2004}]%
        {datar2004locality}
\bibfield{author}{\bibinfo{person}{Mayur Datar}, \bibinfo{person}{Nicole
  Immorlica}, \bibinfo{person}{Piotr Indyk}, {and} \bibinfo{person}{Vahab~S
  Mirrokni}.} \bibinfo{year}{2004}\natexlab{}.
\newblock \showarticletitle{Locality-sensitive hashing scheme based on p-stable
  distributions}. In \bibinfo{booktitle}{\emph{SoCG}}.
  \bibinfo{pages}{253--262}.
\newblock


\bibitem[\protect\citeauthoryear{Deng, Ramachandran, Xu, Hu, Yao, Datta, and
  Zhang}{Deng et~al\mbox{.}}{2025}]%
        {deng2025reforce}
\bibfield{author}{\bibinfo{person}{Minghang Deng}, \bibinfo{person}{Ashwin
  Ramachandran}, \bibinfo{person}{Canwen Xu}, \bibinfo{person}{Lanxiang Hu},
  \bibinfo{person}{Zhewei Yao}, \bibinfo{person}{Anupam Datta}, {and}
  \bibinfo{person}{Hao Zhang}.} \bibinfo{year}{2025}\natexlab{}.
\newblock \showarticletitle{Reforce: A Text-to-SQL agent with self-refinement,
  format restriction, and column exploration}. In
  \bibinfo{booktitle}{\emph{ICLR 2025 Workshop: VerifAI: AI Verification in the
  Wild}}.
\newblock


\bibitem[\protect\citeauthoryear{Dong, Zhang, Ge, Mao, Gao, Lin, Lou,
  et~al\mbox{.}}{Dong et~al\mbox{.}}{2023}]%
        {dong2023c3}
\bibfield{author}{\bibinfo{person}{Xuemei Dong}, \bibinfo{person}{Chao Zhang},
  \bibinfo{person}{Yuhang Ge}, \bibinfo{person}{Yuren Mao},
  \bibinfo{person}{Yunjun Gao}, \bibinfo{person}{Jinshu Lin},
  \bibinfo{person}{Dongfang Lou}, {et~al\mbox{.}}}
  \bibinfo{year}{2023}\natexlab{}.
\newblock \showarticletitle{C3: Zero-shot text-to-sql with chatgpt}.
\newblock \bibinfo{journal}{\emph{arXiv preprint arXiv:2307.07306}}
  (\bibinfo{year}{2023}).
\newblock


\bibitem[\protect\citeauthoryear{Fan, Gu, Zhang, Zhang, Chen, Cao, Li, Madden,
  Du, and Tang}{Fan et~al\mbox{.}}{2024}]%
        {fan2024combining}
\bibfield{author}{\bibinfo{person}{Ju Fan}, \bibinfo{person}{Zihui Gu},
  \bibinfo{person}{Songyue Zhang}, \bibinfo{person}{Yuxin Zhang},
  \bibinfo{person}{Zui Chen}, \bibinfo{person}{Lei Cao},
  \bibinfo{person}{Guoliang Li}, \bibinfo{person}{Samuel Madden},
  \bibinfo{person}{Xiaoyong Du}, {and} \bibinfo{person}{Nan Tang}.}
  \bibinfo{year}{2024}\natexlab{}.
\newblock \showarticletitle{Combining small language models and large language
  models for zero-shot NL2SQL}.
\newblock \bibinfo{journal}{\emph{PVLDB}} \bibinfo{volume}{17},
  \bibinfo{number}{11} (\bibinfo{year}{2024}), \bibinfo{pages}{2750--2763}.
\newblock


\bibitem[\protect\citeauthoryear{Finegan-Dollak, Kummerfeld, Zhang, Ramanathan,
  Sadasivam, Zhang, and Radev}{Finegan-Dollak et~al\mbox{.}}{2018}]%
        {finegan2018improving}
\bibfield{author}{\bibinfo{person}{Catherine Finegan-Dollak},
  \bibinfo{person}{Jonathan~K Kummerfeld}, \bibinfo{person}{Li Zhang},
  \bibinfo{person}{Karthik Ramanathan}, \bibinfo{person}{Sesh Sadasivam},
  \bibinfo{person}{Rui Zhang}, {and} \bibinfo{person}{Dragomir Radev}.}
  \bibinfo{year}{2018}\natexlab{}.
\newblock \showarticletitle{Improving text-to-sql evaluation methodology}.
\newblock \bibinfo{journal}{\emph{arXiv preprint arXiv:1806.09029}}
  (\bibinfo{year}{2018}).
\newblock


\bibitem[\protect\citeauthoryear{Fu, Liu, Wu, Li, Tan, and Sun}{Fu
  et~al\mbox{.}}{2023}]%
        {fu2023catsql}
\bibfield{author}{\bibinfo{person}{Han Fu}, \bibinfo{person}{Chang Liu},
  \bibinfo{person}{Bin Wu}, \bibinfo{person}{Feifei Li}, \bibinfo{person}{Jian
  Tan}, {and} \bibinfo{person}{Jianling Sun}.} \bibinfo{year}{2023}\natexlab{}.
\newblock \showarticletitle{Catsql: Towards real world natural language to sql
  applications}.
\newblock \bibinfo{journal}{\emph{PVLDB}} \bibinfo{volume}{16},
  \bibinfo{number}{6} (\bibinfo{year}{2023}), \bibinfo{pages}{1534--1547}.
\newblock


\bibitem[\protect\citeauthoryear{Gao, Wang, Li, Sun, Qian, Ding, and Zhou}{Gao
  et~al\mbox{.}}{2024b}]%
        {gao2023text}
\bibfield{author}{\bibinfo{person}{Dawei Gao}, \bibinfo{person}{Haibin Wang},
  \bibinfo{person}{Yaliang Li}, \bibinfo{person}{Xiuyu Sun},
  \bibinfo{person}{Yichen Qian}, \bibinfo{person}{Bolin Ding}, {and}
  \bibinfo{person}{Jingren Zhou}.} \bibinfo{year}{2024}\natexlab{b}.
\newblock \showarticletitle{Text-to-sql empowered by large language models: A
  benchmark evaluation}.
\newblock \bibinfo{journal}{\emph{PVLDB}} \bibinfo{volume}{17},
  \bibinfo{number}{5} (\bibinfo{year}{2024}), \bibinfo{pages}{1132--1145}.
\newblock


\bibitem[\protect\citeauthoryear{Gao, Liu, Li, Shi, Zhu, Wang, Li, Li, Hong,
  Luo, et~al\mbox{.}}{Gao et~al\mbox{.}}{2024a}]%
        {gao2024xiyan}
\bibfield{author}{\bibinfo{person}{Yingqi Gao}, \bibinfo{person}{Yifu Liu},
  \bibinfo{person}{Xiaoxia Li}, \bibinfo{person}{Xiaorong Shi},
  \bibinfo{person}{Yin Zhu}, \bibinfo{person}{Yiming Wang},
  \bibinfo{person}{Shiqi Li}, \bibinfo{person}{Wei Li}, \bibinfo{person}{Yuntao
  Hong}, \bibinfo{person}{Zhiling Luo}, {et~al\mbox{.}}}
  \bibinfo{year}{2024}\natexlab{a}.
\newblock \showarticletitle{XiYan-SQL: A Multi-Generator Ensemble Framework for
  Text-to-SQL}.
\newblock \bibinfo{journal}{\emph{arXiv preprint arXiv:2411.08599}}
  (\bibinfo{year}{2024}).
\newblock


\bibitem[\protect\citeauthoryear{Hao, Song, Cai, and Xu}{Hao
  et~al\mbox{.}}{2025}]%
        {hao2025text}
\bibfield{author}{\bibinfo{person}{Zhifeng Hao}, \bibinfo{person}{Qibin Song},
  \bibinfo{person}{Ruichu Cai}, {and} \bibinfo{person}{Boyan Xu}.}
  \bibinfo{year}{2025}\natexlab{}.
\newblock \showarticletitle{Text-to-SQL as Dual-State Reasoning: Integrating
  Adaptive Context and Progressive Generation}.
\newblock \bibinfo{journal}{\emph{arXiv preprint arXiv:2511.21402}}
  (\bibinfo{year}{2025}).
\newblock


\bibitem[\protect\citeauthoryear{Hong, Troynikov, and Huber}{Hong
  et~al\mbox{.}}{2025}]%
        {hong2025context}
\bibfield{author}{\bibinfo{person}{Kelly Hong}, \bibinfo{person}{Anton
  Troynikov}, {and} \bibinfo{person}{Jeff Huber}.}
  \bibinfo{year}{2025}\natexlab{}.
\newblock \bibinfo{booktitle}{\emph{Context rot: How increasing input tokens
  impacts llm performance}}.
\newblock \bibinfo{type}{{T}echnical {R}eport}. \bibinfo{institution}{Chroma}.
\newblock
\urldef\tempurl%
\url{https://trychroma.com/research/context-rot}
\showURL{%
\tempurl}


\bibitem[\protect\citeauthoryear{Lee, Polozov, and Richardson}{Lee
  et~al\mbox{.}}{2021}]%
        {lee2021kaggledbqa}
\bibfield{author}{\bibinfo{person}{Chia-Hsuan Lee}, \bibinfo{person}{Oleksandr
  Polozov}, {and} \bibinfo{person}{Matthew Richardson}.}
  \bibinfo{year}{2021}\natexlab{}.
\newblock \showarticletitle{KaggleDBQA: Realistic evaluation of text-to-SQL
  parsers}. In \bibinfo{booktitle}{\emph{IJCNLP}}. \bibinfo{pages}{2261--2273}.
\newblock


\bibitem[\protect\citeauthoryear{Lee, Park, Kim, and Park}{Lee
  et~al\mbox{.}}{2025}]%
        {lee2025mcs}
\bibfield{author}{\bibinfo{person}{Dongjun Lee}, \bibinfo{person}{Choongwon
  Park}, \bibinfo{person}{Jaehyuk Kim}, {and} \bibinfo{person}{Heesoo Park}.}
  \bibinfo{year}{2025}\natexlab{}.
\newblock \showarticletitle{Mcs-sql: Leveraging multiple prompts and
  multiple-choice selection for text-to-sql generation}. In
  \bibinfo{booktitle}{\emph{COLING}}. \bibinfo{pages}{337--353}.
\newblock


\bibitem[\protect\citeauthoryear{Lei, Chen, Ye, Cao, Shin, Su, Suo, Gao, Hu,
  Yin, et~al\mbox{.}}{Lei et~al\mbox{.}}{2024}]%
        {lei2024spider}
\bibfield{author}{\bibinfo{person}{Fangyu Lei}, \bibinfo{person}{Jixuan Chen},
  \bibinfo{person}{Yuxiao Ye}, \bibinfo{person}{Ruisheng Cao},
  \bibinfo{person}{Dongchan Shin}, \bibinfo{person}{Hongjin Su},
  \bibinfo{person}{Zhaoqing Suo}, \bibinfo{person}{Hongcheng Gao},
  \bibinfo{person}{Wenjing Hu}, \bibinfo{person}{Pengcheng Yin},
  {et~al\mbox{.}}} \bibinfo{year}{2024}\natexlab{}.
\newblock \showarticletitle{Spider 2.0: Evaluating language models on
  real-world enterprise text-to-sql workflows}.
\newblock \bibinfo{journal}{\emph{arXiv preprint arXiv:2411.07763}}
  (\bibinfo{year}{2024}).
\newblock


\bibitem[\protect\citeauthoryear{Lewis, Perez, Piktus, Petroni, Karpukhin,
  Goyal, K{\"u}ttler, Lewis, Yih, Rockt{\"a}schel, et~al\mbox{.}}{Lewis
  et~al\mbox{.}}{2020}]%
        {lewis2020retrieval}
\bibfield{author}{\bibinfo{person}{Patrick Lewis}, \bibinfo{person}{Ethan
  Perez}, \bibinfo{person}{Aleksandra Piktus}, \bibinfo{person}{Fabio Petroni},
  \bibinfo{person}{Vladimir Karpukhin}, \bibinfo{person}{Naman Goyal},
  \bibinfo{person}{Heinrich K{\"u}ttler}, \bibinfo{person}{Mike Lewis},
  \bibinfo{person}{Wen-tau Yih}, \bibinfo{person}{Tim Rockt{\"a}schel},
  {et~al\mbox{.}}} \bibinfo{year}{2020}\natexlab{}.
\newblock \showarticletitle{Retrieval-augmented generation for
  knowledge-intensive nlp tasks}.
\newblock \bibinfo{journal}{\emph{NeurIPS}}  \bibinfo{volume}{33}
  (\bibinfo{year}{2020}), \bibinfo{pages}{9459--9474}.
\newblock


\bibitem[\protect\citeauthoryear{Li, Luo, Chai, Li, and Tang}{Li
  et~al\mbox{.}}{2024b}]%
        {li2024dawn}
\bibfield{author}{\bibinfo{person}{Boyan Li}, \bibinfo{person}{Yuyu Luo},
  \bibinfo{person}{Chengliang Chai}, \bibinfo{person}{Guoliang Li}, {and}
  \bibinfo{person}{Nan Tang}.} \bibinfo{year}{2024}\natexlab{b}.
\newblock \showarticletitle{The dawn of natural language to sql: Are we fully
  ready?}
\newblock \bibinfo{journal}{\emph{PVLDB}} \bibinfo{volume}{17},
  \bibinfo{number}{11} (\bibinfo{date}{July} \bibinfo{year}{2024}),
  \bibinfo{pages}{3318--3331}.
\newblock


\bibitem[\protect\citeauthoryear{Li, Zhang, Fan, Xu, Chen, Tang, and Luo}{Li
  et~al\mbox{.}}{2025}]%
        {li2025alpha}
\bibfield{author}{\bibinfo{person}{Boyan Li}, \bibinfo{person}{Jiayi Zhang},
  \bibinfo{person}{Ju Fan}, \bibinfo{person}{Yanwei Xu}, \bibinfo{person}{Chong
  Chen}, \bibinfo{person}{Nan Tang}, {and} \bibinfo{person}{Yuyu Luo}.}
  \bibinfo{year}{2025}\natexlab{}.
\newblock \showarticletitle{Alpha-sql: Zero-shot text-to-sql using monte carlo
  tree search}.
\newblock \bibinfo{journal}{\emph{arXiv preprint arXiv:2502.17248}}
  (\bibinfo{year}{2025}).
\newblock


\bibitem[\protect\citeauthoryear{Li and Jagadish}{Li and Jagadish}{2014}]%
        {li2014constructing}
\bibfield{author}{\bibinfo{person}{Fei Li} {and} \bibinfo{person}{Hosagrahar~V
  Jagadish}.} \bibinfo{year}{2014}\natexlab{}.
\newblock \showarticletitle{Constructing an interactive natural language
  interface for relational databases}.
\newblock \bibinfo{journal}{\emph{PVLDB}} \bibinfo{volume}{8},
  \bibinfo{number}{1} (\bibinfo{year}{2014}), \bibinfo{pages}{73--84}.
\newblock


\bibitem[\protect\citeauthoryear{Li, Zhang, Li, and Chen}{Li
  et~al\mbox{.}}{2023}]%
        {li2023resdsql}
\bibfield{author}{\bibinfo{person}{Haoyang Li}, \bibinfo{person}{Jing Zhang},
  \bibinfo{person}{Cuiping Li}, {and} \bibinfo{person}{Hong Chen}.}
  \bibinfo{year}{2023}\natexlab{}.
\newblock \showarticletitle{Resdsql: Decoupling schema linking and skeleton
  parsing for text-to-sql}. In \bibinfo{booktitle}{\emph{AAAI}},
  Vol.~\bibinfo{volume}{37}. \bibinfo{pages}{13067--13075}.
\newblock


\bibitem[\protect\citeauthoryear{Li, Zhang, Liu, Fan, Zhang, Zhu, Wei, Pan, Li,
  and Chen}{Li et~al\mbox{.}}{2024c}]%
        {li2024codes}
\bibfield{author}{\bibinfo{person}{Haoyang Li}, \bibinfo{person}{Jing Zhang},
  \bibinfo{person}{Hanbing Liu}, \bibinfo{person}{Ju Fan},
  \bibinfo{person}{Xiaokang Zhang}, \bibinfo{person}{Jun Zhu},
  \bibinfo{person}{Renjie Wei}, \bibinfo{person}{Hongyan Pan},
  \bibinfo{person}{Cuiping Li}, {and} \bibinfo{person}{Hong Chen}.}
  \bibinfo{year}{2024}\natexlab{c}.
\newblock \showarticletitle{Codes: Towards building open-source language models
  for text-to-sql}.
\newblock \bibinfo{journal}{\emph{PACMMOD}} \bibinfo{volume}{2},
  \bibinfo{number}{3} (\bibinfo{year}{2024}), \bibinfo{pages}{1--28}.
\newblock


\bibitem[\protect\citeauthoryear{Li, Hui, Qu, Yang, Li, Li, Wang, Qin, Geng,
  Huo, et~al\mbox{.}}{Li et~al\mbox{.}}{2024a}]%
        {li2024can}
\bibfield{author}{\bibinfo{person}{Jinyang Li}, \bibinfo{person}{Binyuan Hui},
  \bibinfo{person}{Ge Qu}, \bibinfo{person}{Jiaxi Yang},
  \bibinfo{person}{Binhua Li}, \bibinfo{person}{Bowen Li},
  \bibinfo{person}{Bailin Wang}, \bibinfo{person}{Bowen Qin},
  \bibinfo{person}{Ruiying Geng}, \bibinfo{person}{Nan Huo}, {et~al\mbox{.}}}
  \bibinfo{year}{2024}\natexlab{a}.
\newblock \showarticletitle{Can llm already serve as a database interface? a
  Modeling Ambiguityse grounded text-to-sqls}.
\newblock \bibinfo{journal}{\emph{NeurIPS}}  \bibinfo{volume}{36}
  (\bibinfo{year}{2024}), \bibinfo{pages}{42330--42357}.
\newblock


\bibitem[\protect\citeauthoryear{Lin, Asai, Li, Oguz, Lin, Mehdad, Yih, and
  Chen}{Lin et~al\mbox{.}}{2023}]%
        {lin2023train}
\bibfield{author}{\bibinfo{person}{Sheng-Chieh Lin}, \bibinfo{person}{Akari
  Asai}, \bibinfo{person}{Minghan Li}, \bibinfo{person}{Barlas Oguz},
  \bibinfo{person}{Jimmy Lin}, \bibinfo{person}{Yashar Mehdad},
  \bibinfo{person}{Wen-tau Yih}, {and} \bibinfo{person}{Xilun Chen}.}
  \bibinfo{year}{2023}\natexlab{}.
\newblock \showarticletitle{How to train your dragon: Diverse augmentation
  towards generalizable dense retrieval}.
\newblock \bibinfo{journal}{\emph{arXiv preprint arXiv:2302.07452}}
  (\bibinfo{year}{2023}).
\newblock


\bibitem[\protect\citeauthoryear{Liu, Hu, Wen, and Yu}{Liu
  et~al\mbox{.}}{2023}]%
        {liu2023comprehensive}
\bibfield{author}{\bibinfo{person}{Aiwei Liu}, \bibinfo{person}{Xuming Hu},
  \bibinfo{person}{Lijie Wen}, {and} \bibinfo{person}{Philip~S Yu}.}
  \bibinfo{year}{2023}\natexlab{}.
\newblock \showarticletitle{A comprehensive evaluation of ChatGPT's zero-shot
  Text-to-SQL capability}.
\newblock \bibinfo{journal}{\emph{arXiv preprint arXiv:2303.13547}}
  (\bibinfo{year}{2023}).
\newblock


\bibitem[\protect\citeauthoryear{Liu, Shen, Li, Ma, Jiang, Zhang, Fan, Li,
  Tang, and Luo}{Liu et~al\mbox{.}}{2025}]%
        {liu2025survey}
\bibfield{author}{\bibinfo{person}{Xinyu Liu}, \bibinfo{person}{Shuyu Shen},
  \bibinfo{person}{Boyan Li}, \bibinfo{person}{Peixian Ma},
  \bibinfo{person}{Runzhi Jiang}, \bibinfo{person}{Yuxin Zhang},
  \bibinfo{person}{Ju Fan}, \bibinfo{person}{Guoliang Li}, \bibinfo{person}{Nan
  Tang}, {and} \bibinfo{person}{Yuyu Luo}.} \bibinfo{year}{2025}\natexlab{}.
\newblock \showarticletitle{A survey of text-to-sql in the era of llms: Where
  are we, and where are we going?}
\newblock \bibinfo{journal}{\emph{TKDE}} (\bibinfo{year}{2025}).
\newblock


\bibitem[\protect\citeauthoryear{Luo, Li, Fan, Chai, and Tang}{Luo
  et~al\mbox{.}}{2025}]%
        {luo2025natural}
\bibfield{author}{\bibinfo{person}{Yuyu Luo}, \bibinfo{person}{Guoliang Li},
  \bibinfo{person}{Ju Fan}, \bibinfo{person}{Chengliang Chai}, {and}
  \bibinfo{person}{Nan Tang}.} \bibinfo{year}{2025}\natexlab{}.
\newblock \showarticletitle{Natural language to sql: State of the art and open
  problems}.
\newblock \bibinfo{journal}{\emph{PVLDB}} \bibinfo{volume}{18},
  \bibinfo{number}{12} (\bibinfo{year}{2025}), \bibinfo{pages}{5466--5471}.
\newblock


\bibitem[\protect\citeauthoryear{Pourreza, Li, Sun, Chung, Talaei, Kakkar, Gan,
  Saberi, Ozcan, and Arik}{Pourreza et~al\mbox{.}}{2024}]%
        {pourreza2024chase}
\bibfield{author}{\bibinfo{person}{Mohammadreza Pourreza},
  \bibinfo{person}{Hailong Li}, \bibinfo{person}{Ruoxi Sun},
  \bibinfo{person}{Yeounoh Chung}, \bibinfo{person}{Shayan Talaei},
  \bibinfo{person}{Gaurav~Tarlok Kakkar}, \bibinfo{person}{Yu Gan},
  \bibinfo{person}{Amin Saberi}, \bibinfo{person}{Fatma Ozcan}, {and}
  \bibinfo{person}{Sercan~O Arik}.} \bibinfo{year}{2024}\natexlab{}.
\newblock \showarticletitle{Chase-sql: Multi-path reasoning and preference
  optimized candidate selection in text-to-sql}.
\newblock \bibinfo{journal}{\emph{arXiv preprint arXiv:2410.01943}}
  (\bibinfo{year}{2024}).
\newblock


\bibitem[\protect\citeauthoryear{Pourreza and Rafiei}{Pourreza and
  Rafiei}{2023}]%
        {pourreza2023dinsql}
\bibfield{author}{\bibinfo{person}{Mohammadreza Pourreza} {and}
  \bibinfo{person}{Davood Rafiei}.} \bibinfo{year}{2023}\natexlab{}.
\newblock \showarticletitle{{DIN}-{SQL}: Decomposed In-Context Learning of
  Text-to-{SQL} with Self-Correction}. In \bibinfo{booktitle}{\emph{NeurIPS}}.
  \bibinfo{pages}{36339--36348}.
\newblock


\bibitem[\protect\citeauthoryear{Pourreza and Rafiei}{Pourreza and
  Rafiei}{2024}]%
        {pourreza2024dts}
\bibfield{author}{\bibinfo{person}{Mohammadreza Pourreza} {and}
  \bibinfo{person}{Davood Rafiei}.} \bibinfo{year}{2024}\natexlab{}.
\newblock \showarticletitle{Dts-sql: Decomposed text-to-sql with small large
  language models}.
\newblock \bibinfo{journal}{\emph{arXiv preprint arXiv:2402.01117}}
  (\bibinfo{year}{2024}).
\newblock


\bibitem[\protect\citeauthoryear{Qi, Tang, He, Wan, Cheng, Zhou, Wang, Zhang,
  and Lin}{Qi et~al\mbox{.}}{2022}]%
        {qi-etal-2022-rasat}
\bibfield{author}{\bibinfo{person}{Jiexing Qi}, \bibinfo{person}{Jingyao Tang},
  \bibinfo{person}{Ziwei He}, \bibinfo{person}{Xiangpeng Wan},
  \bibinfo{person}{Yu Cheng}, \bibinfo{person}{Chenghu Zhou},
  \bibinfo{person}{Xinbing Wang}, \bibinfo{person}{Quanshi Zhang}, {and}
  \bibinfo{person}{Zhouhan Lin}.} \bibinfo{year}{2022}\natexlab{}.
\newblock \showarticletitle{{RASAT}: Integrating Relational Structures into
  Pretrained {S}eq2{S}eq Model for Text-to-{SQL}}. In
  \bibinfo{booktitle}{\emph{EMNLP}}. \bibinfo{pages}{3215--3229}.
\newblock


\bibitem[\protect\citeauthoryear{Rajkumar, Li, and Bahdanau}{Rajkumar
  et~al\mbox{.}}{2022}]%
        {rajkumar2022evaluating}
\bibfield{author}{\bibinfo{person}{Nitarshan Rajkumar},
  \bibinfo{person}{Raymond Li}, {and} \bibinfo{person}{Dzmitry Bahdanau}.}
  \bibinfo{year}{2022}\natexlab{}.
\newblock \showarticletitle{Evaluating the text-to-sql capabilities of large
  language models}.
\newblock \bibinfo{journal}{\emph{arXiv preprint arXiv:2204.00498}}
  (\bibinfo{year}{2022}).
\newblock


\bibitem[\protect\citeauthoryear{Reimers and Gurevych}{Reimers and
  Gurevych}{2019}]%
        {reimers2019sentence}
\bibfield{author}{\bibinfo{person}{Nils Reimers} {and} \bibinfo{person}{Iryna
  Gurevych}.} \bibinfo{year}{2019}\natexlab{}.
\newblock \showarticletitle{Sentence-bert: Sentence embeddings using siamese
  bert-networks}. In \bibinfo{booktitle}{\emph{EMNLP-IJCNLP}}.
  \bibinfo{pages}{3982--3992}.
\newblock


\bibitem[\protect\citeauthoryear{Singh, Fry, Perelman, Tart, Ganesh, El-Kishky,
  McLaughlin, Low, Ostrow, Ananthram, et~al\mbox{.}}{Singh
  et~al\mbox{.}}{2025}]%
        {singh2025openai}
\bibfield{author}{\bibinfo{person}{Aaditya Singh}, \bibinfo{person}{Adam Fry},
  \bibinfo{person}{Adam Perelman}, \bibinfo{person}{Adam Tart},
  \bibinfo{person}{Adi Ganesh}, \bibinfo{person}{Ahmed El-Kishky},
  \bibinfo{person}{Aidan McLaughlin}, \bibinfo{person}{Aiden Low},
  \bibinfo{person}{AJ Ostrow}, \bibinfo{person}{Akhila Ananthram},
  {et~al\mbox{.}}} \bibinfo{year}{2025}\natexlab{}.
\newblock \showarticletitle{OpenAI GPT-5 System Card}.
\newblock \bibinfo{journal}{\emph{arXiv preprint arXiv:2601.03267}}
  (\bibinfo{year}{2025}).
\newblock


\bibitem[\protect\citeauthoryear{Talaei, Pourreza, Chang, Mirhoseini, and
  Saberi}{Talaei et~al\mbox{.}}{2024}]%
        {talaei2024chess}
\bibfield{author}{\bibinfo{person}{Shayan Talaei},
  \bibinfo{person}{Mohammadreza Pourreza}, \bibinfo{person}{Yu-Chen Chang},
  \bibinfo{person}{Azalia Mirhoseini}, {and} \bibinfo{person}{Amin Saberi}.}
  \bibinfo{year}{2024}\natexlab{}.
\newblock \showarticletitle{CHESS: Contextual Harnessing for Efficient SQL
  Synthesis}.
\newblock \bibinfo{journal}{\emph{arXiv preprint arXiv:2405.16755}}
  (\bibinfo{year}{2024}).
\newblock


\bibitem[\protect\citeauthoryear{Touvron, Martin, Stone, Albert, Almahairi,
  Babaei, Bashlykov, Batra, Bhargava, Bhosale, et~al\mbox{.}}{Touvron
  et~al\mbox{.}}{2023}]%
        {touvron2023llama}
\bibfield{author}{\bibinfo{person}{Hugo Touvron}, \bibinfo{person}{Louis
  Martin}, \bibinfo{person}{Kevin Stone}, \bibinfo{person}{Peter Albert},
  \bibinfo{person}{Amjad Almahairi}, \bibinfo{person}{Yasmine Babaei},
  \bibinfo{person}{Nikolay Bashlykov}, \bibinfo{person}{Soumya Batra},
  \bibinfo{person}{Prajjwal Bhargava}, \bibinfo{person}{Shruti Bhosale},
  {et~al\mbox{.}}} \bibinfo{year}{2023}\natexlab{}.
\newblock \showarticletitle{Llama 2: Open foundation and fine-tuned chat
  models}.
\newblock \bibinfo{journal}{\emph{arXiv preprint arXiv:2307.09288}}
  (\bibinfo{year}{2023}).
\newblock


\bibitem[\protect\citeauthoryear{Wang, Ren, Yang, Liang, Bai, Chai, Yan, Zhang,
  Yin, Sun, et~al\mbox{.}}{Wang et~al\mbox{.}}{2024}]%
        {wang2024mac}
\bibfield{author}{\bibinfo{person}{Bing Wang}, \bibinfo{person}{Changyu Ren},
  \bibinfo{person}{Jian Yang}, \bibinfo{person}{Xinnian Liang},
  \bibinfo{person}{Jiaqi Bai}, \bibinfo{person}{Linzheng Chai},
  \bibinfo{person}{Zhao Yan}, \bibinfo{person}{Qian-Wen Zhang},
  \bibinfo{person}{Di Yin}, \bibinfo{person}{Xing Sun}, {et~al\mbox{.}}}
  \bibinfo{year}{2024}\natexlab{}.
\newblock \showarticletitle{Mac-sql: A multi-agent collaborative framework for
  text-to-sql}.
\newblock \bibinfo{journal}{\emph{arXiv preprint arXiv:2312.11242}}
  (\bibinfo{year}{2024}).
\newblock


\bibitem[\protect\citeauthoryear{Wang, Shin, Liu, Polozov, and Richardson}{Wang
  et~al\mbox{.}}{2020a}]%
        {wang-etal-2020-rat}
\bibfield{author}{\bibinfo{person}{Bailin Wang}, \bibinfo{person}{Richard
  Shin}, \bibinfo{person}{Xiaodong Liu}, \bibinfo{person}{Oleksandr Polozov},
  {and} \bibinfo{person}{Matthew Richardson}.}
  \bibinfo{year}{2020}\natexlab{a}.
\newblock \showarticletitle{{RAT-SQL}: Relation-Aware Schema Encoding and
  Linking for Text-to-{SQL} Parsers}. In \bibinfo{booktitle}{\emph{ACL}}.
  \bibinfo{pages}{7567--7578}.
\newblock


\bibitem[\protect\citeauthoryear{Wang, Yang, Huang, Jiao, Yang, Jiang,
  Majumder, and Wei}{Wang et~al\mbox{.}}{2022}]%
        {wang2022text}
\bibfield{author}{\bibinfo{person}{Liang Wang}, \bibinfo{person}{Nan Yang},
  \bibinfo{person}{Xiaolong Huang}, \bibinfo{person}{Binxing Jiao},
  \bibinfo{person}{Linjun Yang}, \bibinfo{person}{Daxin Jiang},
  \bibinfo{person}{Rangan Majumder}, {and} \bibinfo{person}{Furu Wei}.}
  \bibinfo{year}{2022}\natexlab{}.
\newblock \showarticletitle{Text embeddings by weakly-supervised contrastive
  pre-training}.
\newblock \bibinfo{journal}{\emph{arXiv preprint arXiv:2212.03533}}
  (\bibinfo{year}{2022}).
\newblock


\bibitem[\protect\citeauthoryear{Wang, Wei, Dong, Bao, Yang, and Zhou}{Wang
  et~al\mbox{.}}{2020b}]%
        {wang2020minilm}
\bibfield{author}{\bibinfo{person}{Wenhui Wang}, \bibinfo{person}{Furu Wei},
  \bibinfo{person}{Li Dong}, \bibinfo{person}{Hangbo Bao}, \bibinfo{person}{Nan
  Yang}, {and} \bibinfo{person}{Ming Zhou}.} \bibinfo{year}{2020}\natexlab{b}.
\newblock \showarticletitle{Minilm: Deep self-attention distillation for
  task-agnostic compression of pre-trained transformers}.
\newblock \bibinfo{journal}{\emph{NeurIPS}}  \bibinfo{volume}{33}
  (\bibinfo{year}{2020}), \bibinfo{pages}{5776--5788}.
\newblock


\bibitem[\protect\citeauthoryear{Wang, Zheng, Cao, Zhang, Wei, Fu, Luo, Chen,
  and Bai}{Wang et~al\mbox{.}}{2025}]%
        {wang2025autolink}
\bibfield{author}{\bibinfo{person}{Ziyang Wang}, \bibinfo{person}{Yuanlei
  Zheng}, \bibinfo{person}{Zhenbiao Cao}, \bibinfo{person}{Xiaojin Zhang},
  \bibinfo{person}{Zhongyu Wei}, \bibinfo{person}{Pei Fu},
  \bibinfo{person}{Zhenbo Luo}, \bibinfo{person}{Wei Chen}, {and}
  \bibinfo{person}{Xiang Bai}.} \bibinfo{year}{2025}\natexlab{}.
\newblock \showarticletitle{AutoLink: Autonomous Schema Exploration and
  Expansion for Scalable Schema Linking in Text-to-SQL at Scale}.
\newblock \bibinfo{journal}{\emph{arXiv preprint arXiv:2511.17190}}
  (\bibinfo{year}{2025}).
\newblock


\bibitem[\protect\citeauthoryear{Xu, Ping, Wu, McAfee, Zhu, Liu, Subramanian,
  Bakhturina, Shoeybi, and Catanzaro}{Xu et~al\mbox{.}}{2023}]%
        {xu2023retrieval}
\bibfield{author}{\bibinfo{person}{Peng Xu}, \bibinfo{person}{Wei Ping},
  \bibinfo{person}{Xianchao Wu}, \bibinfo{person}{Lawrence McAfee},
  \bibinfo{person}{Chen Zhu}, \bibinfo{person}{Zihan Liu},
  \bibinfo{person}{Sandeep Subramanian}, \bibinfo{person}{Evelina Bakhturina},
  \bibinfo{person}{Mohammad Shoeybi}, {and} \bibinfo{person}{Bryan Catanzaro}.}
  \bibinfo{year}{2023}\natexlab{}.
\newblock \showarticletitle{Retrieval meets long context large language
  models}.
\newblock \bibinfo{journal}{\emph{arXiv preprint arXiv:2310.03025}}
  (\bibinfo{year}{2023}).
\newblock


\bibitem[\protect\citeauthoryear{Yaghmazadeh, Wang, Dillig, and
  Dillig}{Yaghmazadeh et~al\mbox{.}}{2017}]%
        {yaghmazadeh2017sqlizer}
\bibfield{author}{\bibinfo{person}{Navid Yaghmazadeh}, \bibinfo{person}{Yuepeng
  Wang}, \bibinfo{person}{Isil Dillig}, {and} \bibinfo{person}{Thomas Dillig}.}
  \bibinfo{year}{2017}\natexlab{}.
\newblock \showarticletitle{Sqlizer: query synthesis from natural language}.
\newblock \bibinfo{journal}{\emph{PACMPL}}  \bibinfo{volume}{1}
  (\bibinfo{year}{2017}), \bibinfo{pages}{1--26}.
\newblock


\bibitem[\protect\citeauthoryear{Yu, Zhang, Yang, Yasunaga, Wang, Li, Ma, Li,
  Yao, Roman, Zhang, and Radev}{Yu et~al\mbox{.}}{2018}]%
        {yu-etal-2018-spider}
\bibfield{author}{\bibinfo{person}{Tao Yu}, \bibinfo{person}{Rui Zhang},
  \bibinfo{person}{Kai Yang}, \bibinfo{person}{Michihiro Yasunaga},
  \bibinfo{person}{Dongxu Wang}, \bibinfo{person}{Zifan Li},
  \bibinfo{person}{James Ma}, \bibinfo{person}{Irene Li},
  \bibinfo{person}{Qingning Yao}, \bibinfo{person}{Shanelle Roman},
  \bibinfo{person}{Zilin Zhang}, {and} \bibinfo{person}{Dragomir Radev}.}
  \bibinfo{year}{2018}\natexlab{}.
\newblock \showarticletitle{{S}pider: A Large-Scale Human-Labeled Dataset for
  Complex and Cross-Domain Semantic Parsing and Text-to-{SQL} Task}. In
  \bibinfo{booktitle}{\emph{EMNLP}}. \bibinfo{pages}{3911--3921}.
\newblock


\bibitem[\protect\citeauthoryear{Zhang, Sun, Chen, Pfister, Zhang, and
  Arik}{Zhang et~al\mbox{.}}{2024}]%
        {zhang2024chain}
\bibfield{author}{\bibinfo{person}{Yusen Zhang}, \bibinfo{person}{Ruoxi Sun},
  \bibinfo{person}{Yanfei Chen}, \bibinfo{person}{Tomas Pfister},
  \bibinfo{person}{Rui Zhang}, {and} \bibinfo{person}{Sercan Arik}.}
  \bibinfo{year}{2024}\natexlab{}.
\newblock \showarticletitle{Chain of agents: Large language models
  collaborating on long-context tasks}.
\newblock \bibinfo{journal}{\emph{NeurIPS}}  \bibinfo{volume}{37}
  (\bibinfo{year}{2024}), \bibinfo{pages}{132208--132237}.
\newblock


\bibitem[\protect\citeauthoryear{Zhong, Xiong, and Socher}{Zhong
  et~al\mbox{.}}{2017}]%
        {zhong2017seq2sql}
\bibfield{author}{\bibinfo{person}{Victor Zhong}, \bibinfo{person}{Caiming
  Xiong}, {and} \bibinfo{person}{Richard Socher}.}
  \bibinfo{year}{2017}\natexlab{}.
\newblock \showarticletitle{Seq2sql: Generating structured queries from natural
  language using reinforcement learning}.
\newblock \bibinfo{journal}{\emph{arXiv preprint arXiv:1709.00103}}
  (\bibinfo{year}{2017}).
\newblock


\end{thebibliography}
\end{document}